\documentclass[11pt]{article}

\usepackage[utf8]{inputenc}        
\usepackage[T1]{fontenc}           
\usepackage{lmodern}               
\usepackage{geometry}              
\usepackage{amsmath,amssymb,amsfonts} 
\usepackage{graphicx}              
\usepackage{hyperref}              
\usepackage{xcolor}                
\usepackage{tabularx}
\usepackage{multirow}
\usepackage{authblk}
\usepackage{natbib}                
\usepackage{booktabs}              
\usepackage{siunitx}               
\usepackage{caption}               
\usepackage{subcaption}            
\usepackage{amsmath}
\usepackage{enumitem}              
\usepackage{url}                   

\hypersetup{
    hidelinks=true
}

\title{Global sonde datasets do not support a mesoscale transition in the turbulent energy cascade}

\date{October 23, 2025}

\author[1]{Thomas D. DeWitt\thanks{Email: thomas.dewitt@utah.edu}}
\author[1]{Timothy J. Garrett\thanks{Email: tim.garrett@utah.edu}}
\affil[1]{Department of Atmospheric Sciences, University of Utah}

\begin{document}

\maketitle

\begin{abstract}
Conceptual and theoretical models describing the dynamics of the atmosphere often assume a hierarchy of dynamic regimes, each operating over some limited range of spatial scales. The largest scales are presumed to be governed by quasi-two-dimensional geostrophic turbulence, mesoscale dynamics by gravity waves, and the smallest scales  by 3D isotropic turbulence.  In theory, this hierarchy should  be observable as clear scale breaks in turbulent kinetic energy spectra as one physical mechanism transitions to the next. Here, we show that this view is not supported by global dropsonde and radiosonde datasets of horizontal winds. Instead, the structure function for horizontal wind calculated for vertical separations  between 200\,m and 8\,km has a Hurst exponent of $H_v \approx 0.6$, which is inconsistent with either gravity waves ($H_v = 1)$ or 3D turbulence ($H_v = 1/3)$. For horizontal separations between 200\,km and 1800\,km, the Hurst exponent is $H_h \approx 0.4$, which is inconsistent with quasi-geostrophic dynamics ($H_h = 1)$. We argue that sonde observations are most consistent with a lesser known ``Lovejoy-Schertzer'' model for stratified turbulence where, at all scales, the dynamics of the atmosphere obey a single anisotropic turbulent cascade with  $H_v=3/5$ and $H_h =1/3$. While separation scales smaller than 200 m are not explored here due to measurement limitations, the analysis nonetheless supports a single cohesive theoretical framework for describing atmospheric dynamics, one that might substitute for the more traditional hierarchy of mechanisms that depends on spatial scale.
\end{abstract}

\noindent \textbf{Keywords:} Atmospheric turbulence, stratified turbulence, structure functions, anisotropic scaling, radiosonde observations, mesoscale dynamics

\section{Introduction}
\label{sec:introduction}

The dynamics of Earth's atmosphere are commonly characterized as being governed  by three-dimensional Kolmogorov turbulence at the smallest dynamical scales, gravity waves at the mesoscale, and quasi-geostrophic turbulence at the largest scales \citep{charney1971,gage1986,lindborg1999}. Combined, this ``transition'' paradigm partitions these regimes  according to their respective wavenumber spectra for kinetic energy, where an isotropic -5/3 exponent at small scales (e.g. $E\left(k\right) \propto k^{-5/3}$; \citet{kolmogorov1941})  gives way to a two-dimensional -3 exponent at large scales (e.g. $E\left(k\right) \propto k^{-3}$; \citet{charney1971}) .
\citet{schertzer1985} proposed a very different paradigm that  represents the full range of atmospheric scales, from the  millimeter to the planetary, as being governed by a single theory of anisotropic turbulence. Here, motions are separable according to the direction of the flow, given that the gravitational force acts in only one  direction, rather than the scale of the flow. The spectral exponents  were  theoretically predicted to be $-5/3$ in the horizontal and $-11/5$ in vertical.

The challenge observationally has been that aircraft measurements are most easily performed along isobars where the exponents may differ from their isoheight counterparts. When calculated along isobars, both paradigms predict a transition in the spectral exponent from $-5/3$ to some larger value at scales of hundreds of kilometers. For quasi-geostrophic turbulence, the large-scale isobaric exponent is -3 \citep{charney1971}, while for Lovejoy-Schertzer turbulence, it is -2.4 \citep{lovejoy2009}. 

The question of which paradigm is best supported by observed isobaric spectra has been the subject of considerable debate \citep{lovejoy2009,lindborg2010,frehlich2010b,schertzer2012}. Although there is a consensus that some type of transition is present \citep{nastrom1983,nastrom1984,nastrom1985,gao1998,cho2001,nielsen1967,julian1970,boer1983}, 
a quantitative analysis that includes a statistical  fit to observations is rarely performed, and so it is not possible to conclusively discriminate which theory more closely reflects the true nature of turbulence. Beyond  this isobaric debate, the Lovejoy-Schertzer paradigm has been almost entirely overlooked. Setting aside studies by the authors who originated the idea \citep{lovejoy2007,pinel2012}, the key prediction of the directional dependence of the exponent has not  been tested --  despite it having been  proposed 40 years ago.  


Fundamentally, the dichotomy is between two very different conceptual understandings of how air moves in the atmosphere. The commonly assumed transition paradigm argues  that the dynamics are governed by physics that depends on spatial scale, whereas the Lovejoy-Schertzer paradigm proposes a single dynamical mechanism that governs atmospheric dynamics regardless of spatial scale.  If the atmosphere can be shown to obey Lovejoy-Schertzer scaling, the tantalizing possibility is offered that observations at one scale and a simple  scaling transformation may be all that is necessary to model atmospheric motions at all scales. Determining which paradigm is most closely reflected by observations is the goal of the study presented here.

Because the details are not widely known, in Sect. \ref{sec:theory} we provide an overview of the theory behind Lovejoy-Schertzer turbulence as well as the various scale-dependent alternatives.
Using multiple dropsonde and radiosonde datasets described in Sect. \ref{sec:methods}, we test the theoretical predictions made by both paradigms in Sect. \ref{sec:results}. We consider a wide range of scales ranging from  $200\,\mathrm{m}$ to $20,000\,\mathrm{km}$ with emphasis placed on careful examination of spectra calculated along both the horizontal and vertical directions, the directional dependence being the distinctive prediction of Lovejoy-Schertzer turbulence. 

\section{Theories of isotropic and anisotropic atmospheric turbulence} \label{sec:theory}

In general, turbulence laws serve to  constrain how kinetic energy is distributed across spatial scales. Although these laws are most commonly represented through the power spectrum $E(k)$ of wind velocities as a function of the wavenumber $k$, there are other methods for performing a scale decomposition. Here, we consider the ``structure function'' as it is more robust to irregularly spaced sonde data \citep{lovejoy2007}. The structure function may be thought of as a real space version of the wavenumber spectrum where $k$ is replaced by a separation vector $\mathbf{\Delta r}$ of variable length and direction. Turbulence laws often hold for individual components of the wind vector $\mathbf{v}$, but for simplicity here we only consider the squared magnitude of the vector differences
\begin{align}
    \Delta v^2 \equiv \left(\mathbf{v}(\mathbf{r})-\mathbf{v}(\mathbf{r}+\mathbf{\Delta r})\right)^2. \label{eq:general turbulence law}
\end{align}

Equation \ref{eq:general turbulence law} is very general and it points to two basic questions.\footnote{A third basic question is how do the statistics of $\Delta v^2$ depend on the spatial location $\mathbf{r}$? The turbulence theories  considered in Table \ref{tab:turbulence theories} assume translational invariance, i.e. that any statistics do not depend on location. Translational invariance is unlikely in the atmosphere due to, for example, the altitude and latitude dependence of large scale dynamic features such as the jet stream. Here, we only consider velocity increments averaged over many spatial locations.} First, how do the statistics of $\Delta v^2$ depend on the length of the separation vector $\mathbf{\Delta r}$? Second, how do the statistics of $\Delta v^2$ depend on the direction of $\mathbf{\Delta r}$?
The answer to the second question of separation \emph{direction} is the main subject of this paper. Many of the foundational theories, such as those proposed by Richardson, Kolmogorov, and Obukhov, assume that turbulence statistics are isotropic, or that there is no directional dependence for the statistics of the flow. Before addressing the alternative, anisotropic turbulence, we must first consider how the statistics vary as a function of separation \emph{distance}.

Common to nearly all turbulence laws is the property of ``scale invariance", which requires that  fluctuations, when averaged over many potential realizations of the flow, follow a power-law function with respect to separation distance $\Delta r \equiv |\mathbf{\Delta r}|$ such that
\begin{align}
    \langle\Delta v^2\rangle = \varphi \Delta r^{2H} \label{eq:general structure function}
\end{align}
where $H$ is a constant termed the Hurst exponent and $\varphi$ is some dimensionally relevant quantity that is conserved throughout the turbulent ``cascade'' from one scale to the next.
Note that the kinetic energy spectrum in wavenumber space is obtained from  Eqn. \ref{eq:general structure function} via a Fourier transform to convert the real-space $\Delta x$ into wavenumber $k$ \citep{lovejoy2013}. The kinetic energy spectrum then becomes $E(k)\propto k^{ -(2H+1)}$. 

The next step is to identify the physical quantity $\varphi$ that is conserved during the turbulent cascade. This is the primary aspect by which various theories of turbulence are distinguished. We consider four theories. The first and most widely known was proposed by \citet{kolmogorov1941} where the relevant cascade quantity is the kinetic energy dissipation rate $\varepsilon$, with units of energy per mass per time ($\mathrm{m^2/s^3}$). 
This theory is also known as ``three-dimensional'' turbulence because, in its most basic form, it assumes isotropy, or that the statistics of the flow are identical in all three directions. 

Three-dimensional turbulence has limited relevance for atmospheric motion given that it neglects buoyancy forces. The second theory we consider, proposed independently by \citet{bolgiano1959} and \citet{obukhov1959}, addresses this concern by supposing the conserved cascading quantity relates to buoyancy forces rather than kinetic energy. For a Boussinesq flow, it can be shown that the conserved cascade quantity becomes $\phi=\partial f^2/\partial t$ where $f$ is thermal buoyancy \citep{lovejoy2013}. The quantity $\phi$ is the ``buoyancy variance flux'' with dimensions of acceleration squared per time ($\mathrm{m^2/s^5}$).

Bolgiano and Obukhov's theory was not widely adopted, and it was eventually replaced by various approaches founded on the basic assumption that any feedback between the flow and the stratification is negligible. In these theories,  stratification can influence the flow but the flow cannot modify the stratification--an idea encapsulated by the term ``background'' stratification. To this end, the third theory we consider here was proposed by \citet{charney1971} who adapted the theory of ``two-dimensional'' turbulence \citep{kraichnan1967} to the atmosphere. Traditional two-dimensional turbulence holds that the conserved cascade quantity is ``enstrophy flux'' $\chi$ with units vorticity squared per time ($\mathrm{s^{-3}}$). However, this theory was originally applied only to flows restricted to a plane such as a soap film. Charney modified the theory to account for some limited vertical flow, although the dominant set of dynamics remain a horizontal two-dimensional enstrophy cascade.

The original justification for neglecting feedbacks between the flow and stratification came from gravity wave theory, and so it is not surprising that gravity waves were later adopted directly to explain observed kinetic energy spectra, this time along the vertical direction \citep{vanzandt1982,dewan1997,lindborg2006}. In this fourth theory, the wave frequency, termed the Brunt-V\"ais\"al\"a frequency $N$ with units $\mathrm{s}^{-1}$, takes the place of the conserved cascade quantity. Although $N$ is not a typical cascade quantity, it nonetheless serves as the dimensionally relevant parameter for arguments based on dimensional analysis.

\begin{figure}
    \includegraphics{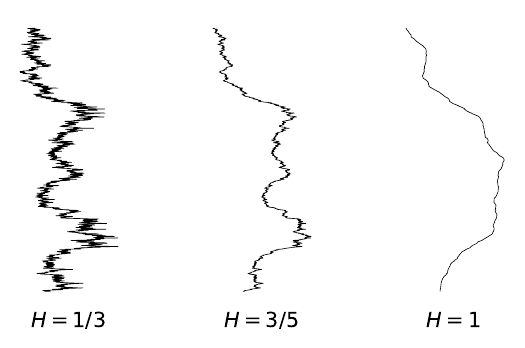}
    \centering \caption{Simulations of a synthetic stochastic process with varying $H$ \citep{lovejoy2010FIF}, representing example wind profiles that might be observed in hypothetical atmospheres where the theories represented by Eqns. \ref{eq:kolmogorov law}-\ref{eq:GW spectrum} apply. The profiles are generated with the same random seed but have varying degrees of ``smoothness'' as specified by the value of $H$.} \label{fig:example multifractals}
\end{figure}

For all the above theories, specification of the relevant cascade quantity is sufficient to determine the value of the Hurst exponent $H$. The value of $H$ is also the main testable prediction that can be used to discriminate between the various theories. Simply by using dimensional analysis,  the kinetic energy fluctuation becomes  a function of the relevant conserved cascade quantity $\varphi$ and the separation distance $r$. In this case, dimensional consistency requires, respectively,
\begin{align}
    &\langle\Delta v^2\rangle = \varepsilon^{2/3} \Delta r^{2H}, \qquad &H=1/3 \qquad &\text{(Kolmogorov spectrum)} \label{eq:kolmogorov law}\\ 
    &\langle\Delta v^2\rangle = \phi^{2/5} \Delta r^{2H}, \qquad &H=3/5 \qquad &\text{(Bolgiano-Obukhov spectrum)} \label{eq:Bolgiano-Obukhov law} \\
    &\langle\Delta v^2\rangle = \chi^{2/3} \Delta r^{2H}, \qquad &H=1 \qquad &\text{(Kraichnan spectrum)}  \label{eq:kraichnan law}\\
    &\langle\Delta v^2\rangle = N^{2} \Delta r^{2H}, \qquad &H=1 \qquad &\text{(Gravity wave spectrum)}  \label{eq:GW spectrum}
\end{align}
Visually, the value of $H$ imparts a unique character to any given transect or profile of $v$. A larger value of $H$ implies a smoother profile, as illustrated in Fig. \ref{fig:example multifractals}.

With respect to the directional dependence of the statistics defined by Eqn. \ref{eq:general structure function}, the simplest case is isotropy, where the statistics are identical for all directions of $\mathbf{\Delta r}$. However, it might instead be expected that the statistics of $\Delta v^2$ vary with direction in the atmosphere given that gravitational stratification has a strong directional dependence. In this case, it is conceivable that multiple theories could hold, even for the same spatial scale. 

One such proposal, first made by \citet{schertzer1985}, considers that gravity, and therefore buoyancy, operate only in the vertical  direction, and so the Bolgiano-Obukhov law (Eqn. \ref{eq:Bolgiano-Obukhov law}) is likely to hold only in the vertical direction. Horizontally, the flow is not bound to a quasi-two dimensional layer as the Kraichnan law requires, so the horizontal statistics are likely to follow the Kolmogorov law (Eqn. \ref{eq:kolmogorov law}):
\begin{align}
    \begin{cases}
    \langle \Delta v^2(\Delta x)\rangle\equiv \langle \Delta v(|\mathbf{\Delta x}|)^2\rangle = \varphi_h \Delta x^{2H_h}, 
    \qquad &\varphi_h  =\varepsilon^{2/3},\qquad H_h=1/3,\\ 
    \langle \Delta v^2(\Delta z)\rangle\equiv \langle \Delta v(|\mathbf{\Delta z}|)^2\rangle = \varphi_v \Delta z^{2H_v}, 
    \qquad &\varphi_v  =\phi^{2/5},\qquad H_v=3/5,
    \end{cases}\label{eq:Lovejoy-Schertzer turbulence, horiz/vert}
\end{align}
where $\mathbf{\Delta x}$ represents a purely horizontal separation vector and $\mathbf{\Delta z}$ represents a purely vertical one. There is no distinction made between the two horizontal directions in this theory, and so  $\mathbf{\Delta x}$ points in any horizontal direction. Together, we term Eqn. \ref{eq:Lovejoy-Schertzer turbulence, horiz/vert} the ``Lovejoy-Schertzer'' theory of turbulence. The two steps of dropping the isotropy assumption and proposing that two different laws hold simultaneously represent a significant shift from how turbulence is normally conceptualized. 

\begin{figure}
    \includegraphics{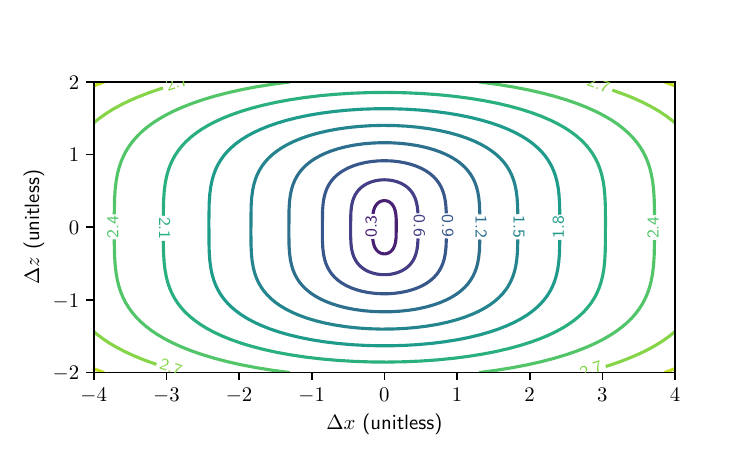}
    \centering \caption{Plot of the Lovejoy-Schertzer turbulent structure function (Eqn. \ref{eq:2D structure function}), representing the average sizes, shapes, and strengths of turbulent circulations. The function is shown in nondimensional form, i.e. $\phi = \varepsilon=1$, and using the theoretical values $H_h=1/3$ and $H_v=3/5$. Note that the empirical structure functions plotted in Sect. \ref{sec:results} consider absolute values for $\Delta x$ and $\Delta z$, which correspond to the first quadrant of this plot.} \label{fig:example 2D structure function}
\end{figure}

It is worth considering the non-obvious implications of this conceptual shift in more detail. One way to view a turbulent velocity field is as a superposition of geometrically simple circulations of varying sizes and wind speeds. The turbulence laws Eqns. \ref{eq:kolmogorov law}-\ref{eq:kraichnan law} may then be interpreted as a relationship between the cross-sectional length and the characteristic velocity of each of these simplified circulations. This picture works because the differences $\Delta v$ used in the structure function effectively isolate the kinetic energy perturbations that are due to circulations of a particular size $\Delta r$.
Thus, isolines of constant $\Delta v^2$ in $\Delta x, \Delta z$ space can be interpreted as describing the shapes and strengths of the simplified circulations from which the overall flow is ``built''. For the special case that turbulence is isotropic, the shapes of the circulations are spherical --  any direction has identical statistics as any other direction. If the turbulence is anisotropic, then the circulations are no longer spherical, as is illustrated in Fig. \ref{fig:example 2D structure function}.

In the Lovejoy-Schertzer theory of turbulence, the isolines are obtained by setting $\langle \Delta v^2(\Delta x)\rangle =\langle \Delta v^2(\Delta z)\rangle $ from Eqn. \ref{eq:Lovejoy-Schertzer turbulence, horiz/vert}. Solving for the aspect ratios of the circulations, we obtain
\begin{align}
    \frac{\Delta x}{\Delta z} = l_s^{-4/5}\Delta z^{4/5};\qquad l_s=\frac{\varepsilon^{5/4}}{\phi^{3/4}}. \label{eq:spheroscale}
\end{align}
Equation \ref{eq:spheroscale} is the first nontrivial implication of the Lovejoy-Schertzer theory. It implies that the mean aspect ratio of circulations systematically changes with circulation size (Fig. \ref{fig:example 2D structure function}). Indeed, this scaling of aspect ratio with size aligns with our intuitive notion that large-scale atmospheric circulations such as the Hadley cell are highly elongated in the horizontal direction, whereas small-scale circulations such as small convective elements are more spherical or even elongated vertically. 
Because the aspect ratios systematically change with scale, there is a unique scale for which the aspect ratio is equal to unity and circulations are nearly spherical, termed the ``spheroscale'' $l_s$ \citep{lovejoy1985,schertzer1985b}.

Another way to examine the shape of the turbulent circulations is to consider the full two-dimensional structure function $\langle \Delta v(\Delta x, \Delta z)^2\rangle$. The simple requirement that $\langle \Delta v(\Delta x, \Delta z)^2\rangle$ reduces to Eqns. \ref{eq:Lovejoy-Schertzer turbulence, horiz/vert} when $\Delta x = 0$ or $\Delta z=0$ is not enough to specify a unique function. If we further require that isotropic turbulence is recovered when $H_h=H_v$ and $\varphi_v=\varphi_h$, then a unique function is specified, as shown  in Fig. \ref{fig:example 2D structure function}:
\begin{align}
    \langle \Delta v(\Delta x, \Delta z)^2\rangle  = \left(\varphi_h^{1/H_h} \Delta x^2  +\varphi_v^{1/H_h} \Delta z^{2H_v/H_h}\right)^{H_h}. \label{eq:2D structure function}
\end{align}

We now fit the two-dimensional structure function given by Eqn. \ref{eq:2D structure function} to observed wind statistics, allowing for both exponents $H_v$ and $H_h$, as well as the coefficients $\varphi_v$ and $\varphi_h$, to be determined empirically. With these four free parameters, Eqn. \ref{eq:2D structure function} becomes very general. It includes as special cases each of the theories for turbulence mentioned so far. An additional case worth mentioning is that the exponents are equal but the constants $\varphi_h$ and $\varphi_v$  have different values. In this case, circulations have non-unitary aspect ratios but the aspect ratio does not change with circulation size. This type of anisotropy has been termed ``trivial anisotropy'' by \citet{lovejoy2013}. Studies investigating turbulent anisotropy based on the anisotropy stress tensor limit themselves to trivial anisotropy if they assume the cascade remains controlled by kinetic energy with $H_h=H_v$, as is sometimes done \citep{tennekes1972}. Table \ref{tab:turbulence theories} summarizes how Eqn. \ref{eq:2D structure function} relates to various turbulent theories.

\begin{table}
    \centering
    \renewcommand{\arraystretch}{1.2} 
    \centering \caption{Various theories of atmospheric turbulence and corresponding parameter values for Eqn. \ref{eq:2D structure function}. Here, $N$ is the Brunt-V\"ais\"al\"a frequency, $\chi$ enstrophy flux, $\phi$ buoyancy variance flux, and $\varepsilon$ kinetic energy flux. ``Undefined'' means that the parameter has no meaning within the context of the theory, while ``not specified'' indicates the parameter has some value but it is not specified by the theory.
    Note that Lindborg's theory \citep{lindborg2006} suggests a transition between $H_v=1$ to $H_v=1/3$ at the Ozmidov length scale of order $3\,\mathrm{m}$. Given that our measurements are at larger scales, we only consider the value $H_v=1$.}\label{tab:turbulence theories}
    \begin{tabularx}{\textwidth}{l l X X}
        \hline
        \textbf{Type} & \textbf{Case} & \textbf{Exponents} & \textbf{Constants} \\
        \hline
        \multirow{3}{*}{Isotropic} & 3D \citep{kolmogorov1941} & $H_h = H_v = 1/3$ & $\varphi_h = \varphi_v=\varepsilon^{2/3}$ \\
                                   & \citet{bolgiano1959,obukhov1959} & $H_h = H_v = 3/5$ & $\varphi_h = \varphi_v=\phi^{2/5}$ \\
                                   & 2D \citep{kraichnan1967} & $H_h = 1$, $H_v$ undefined & $\varphi_h=\chi^{2/3}$, $ \varphi_v$ undefined \\
                                   & Gravity waves \citep{vanzandt1982} & $H_v=1$, $H_h$ not specified & $\varphi_v=N^2$, $\varphi_h$ not specified \\
        \hline
        \multirow{3}{*}{Anisotropic} 
                                    & Quasi-Geostrophic \citep{charney1971}& $H_h = H_v=1$ & $\varphi_h =\chi_h^{2/3}, \varphi_v=\chi_v^{2/3}$, $\chi_v\ll \chi_h$ \\
                                    & \citet{schertzer1985} & $H_h=1/3, H_v=3/5$ & $\varphi_h=\varepsilon^{2/3}, \varphi_v=\phi^{2/5}$ \\
                                    & \citet{lindborg2006} & $H_h=1/3, H_v=1$ & $\varphi_h=\varepsilon^{2/3}, \varphi_v=N^2$ \\
                                    & Anisotropic 3D & $H_h = H_v=1/3$ & $\varphi_h =\varepsilon_h^{2/3}, \varphi_v=\varepsilon_v^{2/3}$, $\varepsilon_h\ne \varepsilon_v$ \\
        \hline
    \end{tabularx}
\end{table}

\section{Methods} \label{sec:methods}

We calculate structure functions from three datasets of wind velocity. The first contains dropsonde measurements from the ACTIVATE field campaign \citep{vomel2023}, which took place over the North Atlantic Ocean between 2020 and 2022. Drops occurred during 169 flights spread over a variety of meteorological conditions and seasons.
The second dropsonde dataset considered here was obtained from NOAA hurricane reconnaissance flights that took place between 1996 and 2012, mainly over the Gulf of Mexico and the Atlantic Ocean.
Profiles were not considered if the data quality was marked as degraded or if the profile did not span the entire layer considered. We analyze a total of 683 ACTIVATE and 2325 hurricane profiles. Both dropsonde datasets contain vertical wind profiles $w(z)$ derived from the measured fall speed, but the uncertainty in the measurements is of order $\sim 1\,\mathrm{m/s}$ \citep{wang2015,vomel2023} -- too large for the purposes of calculating a structure function. For this reason, we only consider structure functions calculated using horizontal vector differences $\Delta v \equiv |\mathbf{v_h}(\mathbf{r})-\mathbf{v_h}(\mathbf{r}+\mathbf{\Delta r})|$ where $\mathbf{v_h}$ is the horizontal component of the full wind vector $\mathbf{v}$.

The third dataset we consider is the horizontal wind data from the Integrated Global Radiosonde Archive (IGRA) \citep{durre2006,durre2018}, a composite created from global  balloon-borne soundings obtained from thousands of stations and spanning many decades. 
A limitation of the dataset is that measurement methods and techniques vary temporally and spatially. For example, many modern sensors use GPS to measure geopotential height, but height inferred from pressure is also used, particularly for older data. 
Accordingly, we only consider measurements from the period between the years 2010 and 2025 when GPS data can be assumed to be in sufficiently widespread use. Measurement uncertainties for the two models of radiosonde widely used during this period, the Vaisala RS41 \citep{vaisala_rs41_docs,dirksen2014}
and the Graw DFM-17, \citep{graw_dfm17_docs, dirksen2014} 
are of order $10\,\mathrm{m}$ (geopotential height), $1\,\mathrm{hPa}$ (pressure), and $0.1\,\mathrm{m/s}$ (horizontal wind). 
The variety of measurement techniques and sensors used to compose the IGRA dataset introduces a source of uncertainty that is difficult to quantify, given that the techniques are not reported in the dataset. As an example, in Appendix \ref{sec:individual stations} we show a near perfect split in calculated Hurst exponents based on sounding nationality, a result likely caused by processing techniques being standard within a country but differing between countries. Any uncertainties we report below should therefore be interpreted with significant caution. 

\begin{figure}
    \includegraphics[width=.5\linewidth]{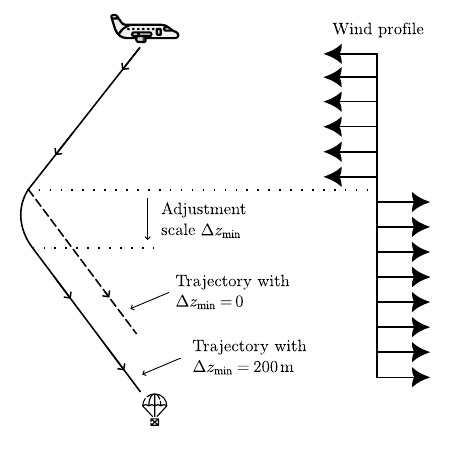}
    \centering \caption{Illustration of how dropsonde measurements of horizontal wind fluctuations $\Delta v(\Delta z)$ are effectively smoothed by sonde inertia. Wind fluctuations that occur over a smaller spatial scale than the sonde's adjustment scale $\Delta z_\text{min}$ cannot be reliably measured.} \label{fig:sonde inertia sketch}
\end{figure}

Shear measurements calculated from radiosonde and dropsonde horizontal wind profiles also have uncertainties related to sonde inertia (Fig. \ref{fig:sonde inertia sketch}). If a sonde passes between layers with different mean wind speeds, the sonde does not immediately adjust to the new wind speed due to its inertia. Since horizontal wind velocity is approximated by the velocity of the sonde itself, shear cannot be measured over any distance smaller than the distance over which the sonde adjusts to a different horizontal wind. To account for this adjustment scale, and to remove additional spurious high-frequency wind variability due to e.g. oscillations caused by the payload swinging underneath the balloon, a smoothing is commonly applied to the wind profiles during data processing. The smoothing applied to ACTIVATE and hurricane sondes had a timescale of $5\,\mathrm{s}$ \citep{vomel2023} 
and $10\,\mathrm{s}$, \citep{durre2006}   
respectively, while a typical radiosonde smoothing timescale is $40\,\mathrm{s}$ \citep{dirksen2014}.
Therefore, assuming typical descent/ascent rates of $20\,\mathrm{m/s}$ (dropsondes; \citet{vomel2023, wang2015}) and $5\,\mathrm{m/s}$ (radiosondes; \citet{dirksen2014}), the spatial scales for sonde adjustment  are of order $200\,\mathrm{m}$ in both cases. We therefore limit our analysis of vertical wind fluctuations to scales larger than $\Delta z_\text{min} = 200\,\mathrm{m}$.

To obtain $H$ in Eqns. \ref{eq:general structure function} and \ref{eq:2D structure function}, only second order structure functions are calculated here, although structure functions of other orders have also been considered by other studies. For comparison with prior results, we reproduce in the Supplement our analysis for first- and third-order structure functions $\langle \Delta v \rangle$ and $\langle \Delta v^3 \rangle$, respectively. All structure functions are calculated from all possible point pairs within a given profile, time period, or region, and fitted values are obtained using a least-squares regression. Uncertainties are reported as 95\% confidence intervals for the least-squares fit and do not include systematic bias originating from processing methodologies such as dataset smoothing.

For the IGRA data set, the structure functions are calculated for both vertical and horizontal separation directions. Purely vertical structure functions are calculated from individual sondes, each released for the 00z and 12z launch times between 2010 and 2025. Structure functions calculated along the horizontal direction are obtained using observations from different devices, first by identifying nearly simultaneous sonde launches. Then, each point observation is paired with each other point observation to obtain a list of observation pairs with varying horizontal and vertical separations. To be included in the analysis, observation pairs are required to have taken place within 2 hours of each other. For one-dimensional structure functions calculated along the horizontal direction,  observations are also required to have vertical separations no larger than $50\,\mathrm{m}$.
Due to the volume of data, it is not possible to consider all observation pairs from all possible sounding launches, and so  only soundings launched at 00z or 12z from every tenth day between 2010 and 2025 are considered for any structure function calculated along the horizontal direction.  

\subsection{Effect of non-instantaneous sonde measurements}

Here, we ignore temporal variability in the statistics for turbulent wind fluctuations in order to isolate the spatial statistics as given by Eqn. \ref{eq:Lovejoy-Schertzer turbulence, horiz/vert}. However, the sondes do not in fact measure wind profiles instantaneously. Dropsondes obtain profiles over approximately one hour, while radiosondes require two to three hours.

To estimate whether time differences between observation pair measurements impact our analyses, consider that, for isotropic turbulence, a turbulent circulation of size $l$ has a lifetime of order $\tau \sim l/\Delta v$ where $\Delta v$ is the wind speed associated with the circulation. For the spatial statistics of the circulation to be accurately sampled, the sonde must cross the circulation within a time interval that is shorter than the circulation lifetime. For sonde velocity $V$, this requires $l/V < \tau$ or equivalently $V>\Delta v$, implying that the sonde can only measure turbulent velocity fluctuations of a magnitude smaller than the sonde velocity. Given that  dropsondes with vertical velocities between 15 and 20 $\mathrm{m/s}$ satisfy this condition for most of the measured range of  scales, and that their measurements also support anisotropy, this isotropic criterion is less relevant.

For an anisotropic circulation, the horizontal and vertical circulation sizes are not identical, and the circulation's lifetime must be estimated using its horizontal length given that we are considering horizontal velocity components. In this case, accurate measurements of the vertical profile require $V>v l_z/l_x$ where $l_z$ is the vertical circulation length or the distance the sonde must pass through to sample a single circulation. From Eqn. \ref{eq:spheroscale}, large eddies are most anisotropic with aspect ratios $l_z/l_x\ll 1$. The largest measured velocity difference among spatially separated measurements in the dataset is  $\sim 20\,\mathrm{m/s}$, associated with eddies that are the most stratified with the smallest values of $l_z/l_x$ according to Eqn. \ref{eq:spheroscale}. These velocity differences are nonetheless only approximately four times faster than the slowest sonde velocity of $5\,\mathrm{m/s}$ (for the radiosondes). The sonde measurements may therefore be assumed effectively instantaneous.

We also include in our analysis horizontally separated measurement pairs that are separated in time by up to two hours. 
A typical circulation of horizontal size $l_x$ has a typical lifetime of $\tau = l_x/\Delta v$ where $\Delta v$ is the wind speed associated with the circulation. The minimum distance in our horizontal measurement pairs is set to $2\times 10^5\,\mathrm{m}$, so that the minimum circulation lifetime is of order $\tau = 2\times 10^5\,\mathrm{m} /10\,\mathrm{m/s} = 2\times 10^4\,\mathrm{s}$, or about 5.5 hours, implying that observation pairs separated by a maximum of two hours may be considered effectively instantaneous. 
As a final check that our thresholds are sufficient, we also computed a one-dimensional horizontal structure function for observation pairs that had vertical separations less than $5\,\mathrm{m}$ and were taken within $5\,\mathrm{min}$ of each other (see Supplement S4). Results for both sets of thresholds ($5\,\mathrm{m}$, $5\,\mathrm{min}$) and ($50\,\mathrm{m}$, $120\,\mathrm{min}$) indicated Hurst exponents that were nearly identical.

\section{Results} \label{sec:results}

\begin{figure}
    \includegraphics{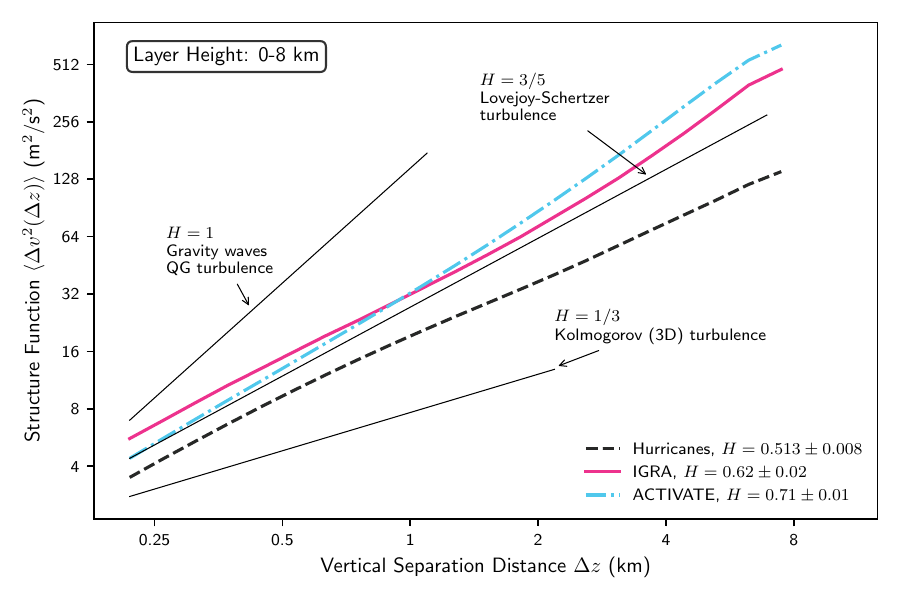}
    \centering \caption{Vertical structure functions for the IGRA radiosonde dataset (pink solid), the NOAA hurricane dropsonde dataset (black dashed) and the ACTIVATE dropsonde dataset (blue dot-dashed). Structure functions and Hurst exponents (Eqn. \ref{eq:general structure function}) are calculated for the lowest 8km of the troposphere, which is the layer measured by all three datasets. }\label{fig:8km vertical spectrum}
\end{figure}

First, we evaluate the structure function along the vertical direction (Eqn. \ref{eq:Lovejoy-Schertzer turbulence, horiz/vert})  for the lowest 8\,km of the atmosphere, which is the layer measured by both dropsonde datasets and IGRA radiosondes. As shown in Fig. \ref{fig:8km vertical spectrum}, the structure functions closely follow a power-law relationship over the full range of observed scales $0.2\,\mathrm{km} \le \Delta z \le 8\,\mathrm{km}$. Calculated values of $H_v$ range from $0.513 \pm 0.008$ for the hurricane dataset to $0.71 \pm 0.01$ for ACTIVATE. 

\begin{figure}
    \includegraphics{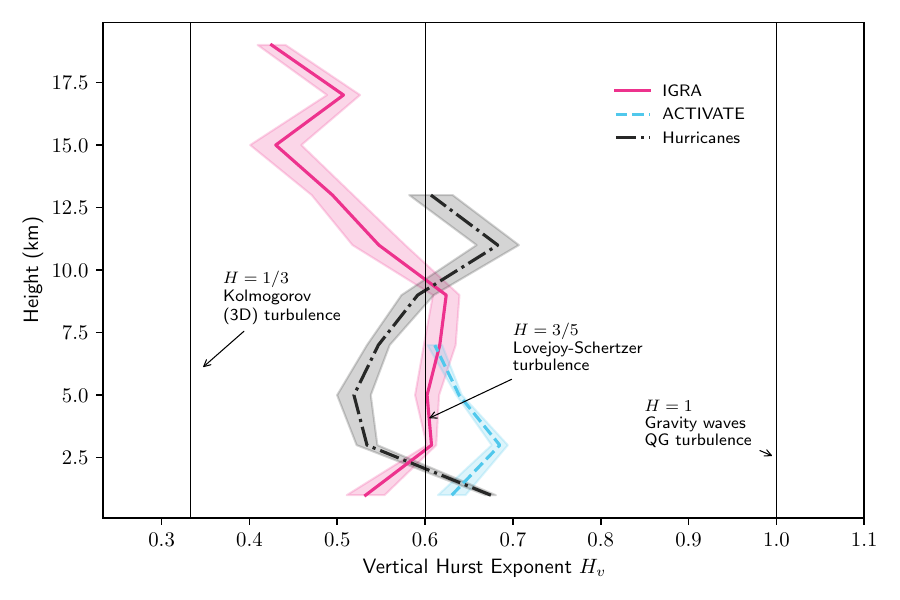}
    \centering \caption{Hurst exponents and 95\% confidence (shaded) calculated for structure functions as shown in Fig. \ref{fig:8km vertical spectrum} but evaluated within  stacked layers $2\,\mathrm{km}$ thick. }\label{fig:exponents with height}
\end{figure}

To investigate the dependence of the vertical Hurst exponent with height, wind measurements are divided into layers of thickness $2\,\mathrm{km}$. A Hurst exponent $H_v$ is calculated for the vertical structure function for $\Delta z$ ranging between $0.2\,\mathrm{km}$ and $2\,\mathrm{km}$ for each layer. Fig. \ref{fig:exponents with height} highlights how the calculated Hurst exponents lie close to the Lovejoy-Schertzer predicted value at each level within the troposphere. The structure functions for individual layers that are used to construct Fig. \ref{fig:exponents with height} are provided in the Supplement.  Above an altitude of roughly 18 to $20\,\mathrm{km}$, $H_v$ decreases to a value lying between approximately 0.4 and 0.5, which does not precisely agree with  any of the theories for turbulence described in Tab. \ref{tab:turbulence theories}.  As shown in the Supplement, tropospheric Hurst exponents for first- and third-order structure functions ($\langle \Delta v\rangle$ and $\langle \Delta v^3 \rangle$, respectively) are also consistent with the Lovejoy-Schertzer prediction.

\begin{figure}
    \includegraphics{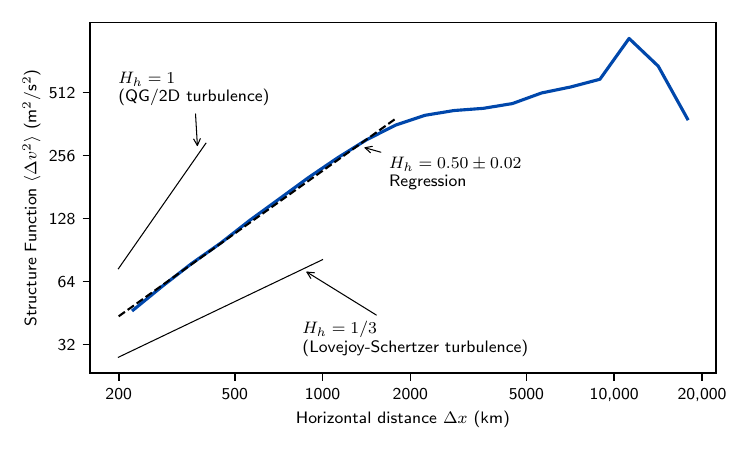}
    \centering \caption{Horizontal structure function $\langle \Delta v (\Delta x)^2\rangle$ calculated from IGRA radiosonde data and the associated theoretical power-law relationships for different turbulence theories shown in  Table \ref{tab:turbulence theories}. }\label{fig:horizontal structure function}
\end{figure}

Horizontal structure functions $\langle \Delta v (\Delta x)^2\rangle$ calculated using IGRA data are shown in Fig. \ref{fig:horizontal structure function}. There data show clear power-law behavior for separations smaller than about $1800\,\mathrm{km}$, while for larger scales the Hurst exponent approaches 0 between approximately $3000\,\mathrm{km}$ and the planetary half-circumference of $20000\,\mathrm{km}$. Regressions to the steepest portion of the slope between $200\,\mathrm{km}$ and $1800\,\mathrm{km}$ indicate a value for $H_h$ of $0.50 \pm 0.02$. 

\subsection{Two-dimensional structure functions}

That the values of the vertical and horizontal Hurst exponents are different supports the view that atmospheric turbulence is non-trivially anisotropic, or that the aspect ratios of atmospheric circulations may systematically change with scale as implied by Eqn. \ref{eq:spheroscale}. To address this possibility, we now examine in detail the full two-dimensional structure functions represented by Eqn. \ref{eq:2D structure function}. 
Note that Eqn. \ref{eq:2D structure function} cannot be easily logarithmically transformed, so what follows is limited to analyses calculated in linear space.

\begin{figure}
    \includegraphics{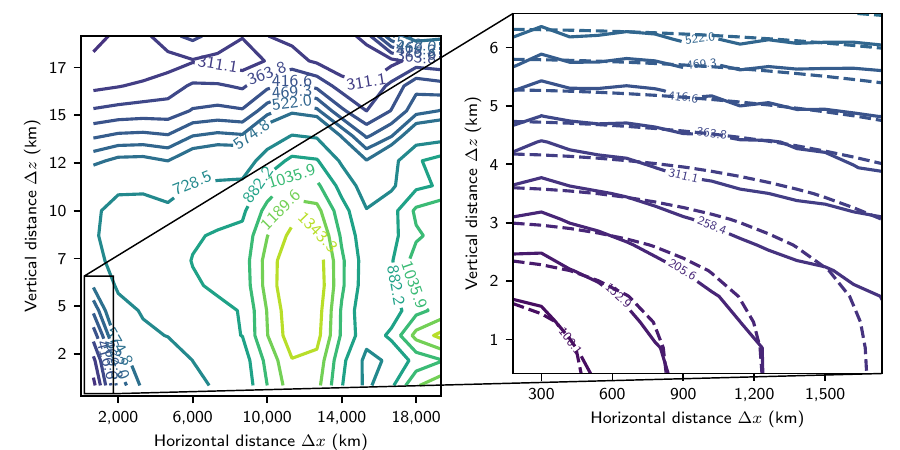}
    \centering \caption{Empirical two-dimensional horizontal/vertical structure function calculated from IGRA radiosonde data (solid; labels in $\mathrm{m^2 s^{-2}}$). For the inset, a fit (dashed) was performed using Eqn. \ref{eq:2D structure function}, with best-fit parameters described in text.}\label{fig:2Dvertical-horizontal structure function}
\end{figure}

Fig. \ref{fig:2Dvertical-horizontal structure function} shows the empirical two-dimensional structure function $\langle \Delta v(\Delta x, \Delta z)^2\rangle $ plotted for two ranges of scale. As was also seen in Fig. \ref{fig:horizontal structure function},  the largest scales up to $20000\,\mathrm{km}$ horizontally and $20\,\mathrm{km}$ vertically show no clear power-law dependence of $\Delta v^2$ on $\Delta x$ or $\Delta z$, indicative of a regime that is not dominated by any turbulence theory listed in Table \ref{tab:turbulence theories}. By contrast, for horizontal separations between $200\,\mathrm{km}$ and $1800\,\mathrm{km}$, and for vertical separations smaller than $7\,\mathrm{km}$, the empirical structure function is well approximated by a least-squares fit to Eqn. \ref{eq:2D structure function}, with empirical values $H_h = 0.37 \pm 0.01$, $H_v = 0.63 \pm 0.01$, $\varphi_h=0.006 \pm 0.002 \,\mathrm{m^{2-2H_h} s^{-2}}$, and $\varphi_v=0.009 \pm 0.002\,\mathrm{m^{2-2H_v} s^{-2}}$. 

From Eqn. \ref{eq:spheroscale} and the theoretical relations $\varphi_h = \varepsilon^{2/3}$ and $\varphi_v = \phi^{2/5}$ (Tab. \ref{tab:turbulence theories}), empirical values for $\varphi_h$ and $\varphi_v$ imply a spheroscale of order $1\mathrm{m}$. Values calculated for $H_h$ and $H_v$ are nearly consistent whether they are calculated from the two-dimensional structure functions shown in Fig. \ref{fig:2Dvertical-horizontal structure function} or from the one-dimensional structure functions shown in Figs. \ref{fig:8km vertical spectrum}-\ref{fig:horizontal structure function}.

\begin{figure}
    \includegraphics{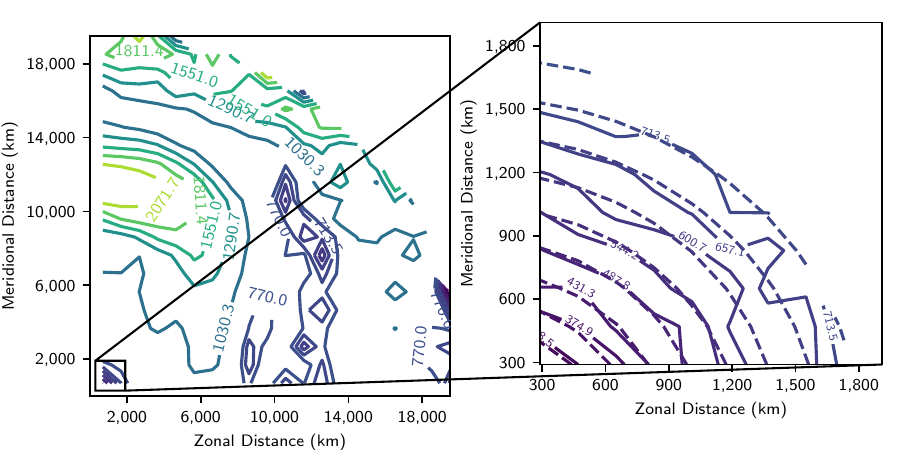}
    \centering \caption{As in Fig. \ref{fig:2Dvertical-horizontal structure function}, but for isoheight statistics calculated as a function of meridional and zonal direction. }\label{fig:2D horizontal structure function}
\end{figure}

Although we are unaware of any theory that predicts different Hurst exponents for the two horizontal directions, for completeness an isoheight two-dimensional structure function for the zonal ($x$) and meridional ($y$) directions is calculated from the sounding observations as shown in Fig. \ref{fig:2D horizontal structure function}. 
The structure function displays a maximum in $\Delta v^2$ near $\Delta x\sim 0$ and $\Delta y \sim 10,000\,\mathrm{km}$, which might be speculated to correspond to the jet stream. Otherwise, there is a ``flattening'' with $H_h\to 0$ at horizontal separation scales larger than $\sim 1800\,\mathrm{km}$ as in Fig. \ref{fig:horizontal structure function}. 

For separations smaller than $1800\,\mathrm{km}$, the empirical structure function is fit to a functional form that is analogous to Eqn. \ref{eq:2D structure function} but that applies in the $x$ and $y$ two horizontal directions:
\begin{align}
    \langle\Delta v(\Delta x, \Delta y)^2\rangle = \left(\varphi_x^{1/H_x} \Delta x^2  +\varphi_y^{1/H_x} \Delta y^{2H_y/H_x}\right)^{H_x}. \label{eq:2D h-h structure function}
\end{align}
Values for the least-squares fit  are $H_x = 0.35 \pm 0.05$, $H_y = 0.33 \pm 0.03$, $\varphi_x=0.03 \pm 0.04\,\mathrm{m^{2-2H_x} s^{-2}}$, and $\varphi_y=0.05 \pm 0.04\,\mathrm{m^{2-2H_y} s^{-2}}$. These values are consistent with the turbulence being either horizontally isotropic or ``trivially anisotropic'', with $H_x\simeq H_y$ but $\varphi_x\ne \varphi_y$, as noted also by \citet{lovejoy2011} based on an examination of reanalysis datasets. 
The mean aspect ratio of the circulations is poorly constrained when fitting all four parameters simultaneously, but constraining the exponents to the theoretical value $H_x=H_y=1/3$ yields $\varphi_x/\varphi_y = 0.90\pm 0.03$.

\section{Discussion} \label{sec:discussion}

For structure functions calculated along the vertical direction, the prediction of the  prevailing ``transition'' paradigm is that $H_v=1$ for the largest vertical scales, whether the model is a quasi-two-dimensional enstrophy cascade \citep{charney1971} or gravity waves \citep{dewan1997}. At smaller scales there is a transition to Kolmogorov turbulence where $H_v=1/3$. Neither value of $H_v$ is supported by Figs. \ref{fig:8km vertical spectrum} and \ref{fig:exponents with height}, at least for vertical separations down to $200\,\mathrm{m}$.

It is especially notable that the Kolmogorov value of $H_v=1/3$ is not supported even within the boundary layer below $2\,\mathrm{km}$ in altitude (Fig. \ref{fig:exponents with height}). This challenges the prevailing viewpoint that Kolmogorov turbulence, in either its isotropic or anisotropic forms (Table \ref{tab:turbulence theories}), characterizes  kilometer-scale boundary layer turbulence. If Kolmogorov turbulence does apply to the boundary layer,  Fig. \ref{fig:exponents with height} suggests that it can only exist for vertical separation scales smaller than $200\,\mathrm{m}$ that are not resolved here.

For large-scale structure functions calculated along the horizontal direction (Fig. \ref{fig:horizontal structure function}), the calculated value of $H_h=0.50\pm 0.02$ is a little higher than the $H_h=1/3$ value predicted by Lovejoy-Schertzer turbulence. Nonetheless, it lies closer to 1/3 than the value of $H_h=1$ expected for a two-dimensional turbulent enstrophy cascade as suggested by \citet{charney1971} and \citet{nastrom1983}, and therefore these results also appear to invalidate the transition paradigm. 

Overall, Hurst exponents for both the horizontal and vertical separation directions appear more strongly supportive of the Lovejoy-Schertzer paradigm of stratified turbulence (Eqn. \ref{eq:Lovejoy-Schertzer turbulence, horiz/vert}) than the transition paradigm. The two dimensional structure function $\Delta v (\Delta x, \Delta z)^2$ (Fig. \ref{fig:2Dvertical-horizontal structure function})  provides what is perhaps the most compelling support as it is not restricted to the orthogonal horizontal and vertical directions and therefore was calculated using many more observation pairs. A fit using Eqn. \ref{fig:2Dvertical-horizontal structure function} approximately reproduces the empirical isolines of constant $\Delta v$, and the best-fit exponent values $H_h=0.37 \pm 0.01$, $H_v=0.63 \pm 0.01$ are again
close to the theoretical values predicted by the Lovejoy-Schertzer theory of turbulence of $H_h=1/3$ and $H_v=3/5$. 

\subsection{The effect of vertical smoothing on the Hurst exponents} \label{sec:vertical exponent smoothing}

The largest discrepancy between the  empirically derived exponents obtained here and those predicted by the Lovejoy-Schertzer theory of turbulence is the value $H_h=0.50\pm 0.02$ obtained from the one-dimensional horizontal structure function in Fig. \ref{fig:horizontal structure function}. In fact, it does not clearly match any of the turbulence theories shown in Table \ref{tab:turbulence theories}. 

The discrepancy is reminiscent of the isobaric spectrum controversy discussed in the introduction -- even if $H = 1/3$ along isoheights it is possible that $H>1/3$ along isobars.
\citet{lovejoy2009}  argued that the basic reason that isobaric and isoheight structure functions differ is that isobars gently slope over large horizontal distances. As an example consider a sloping trajectory where $\Delta z = c \Delta x$ and $c\ll 1$. In this case, from Eqn. \ref{eq:2D structure function}, the observed structure function would follow
\begin{align}
    \langle\Delta v(\Delta x)^2\rangle = \left(\varphi_h^{1/H_h} \Delta x^2  +\varphi_v^{1/H_h} (c \Delta x)^{2H_v/H_h}\right)^{H_h}.\label{eq:sloping 2D structure function}
\end{align}
Given that $H_v/H_h>1$ (Fig. \ref{fig:2Dvertical-horizontal structure function}), at small scales the $\Delta x^{2}$ term dominates so that the observed structure function scales as $\langle\Delta v(\Delta x)^2\rangle \sim \Delta x^{2H_h}$, while at larger scales the $\Delta x^{2H_y/H_x}$ term dominates so that $\langle\Delta v(\Delta x)^2\rangle \sim \Delta x^{2H_v}$.
The implication is that a sloping trajectory will transition from $H\approx 1/3$ at smaller scales to $H\approx 3/5$ at some scale depending on the value of $c$. Such a transition is consistent with the well-known spectrum observed by \citet{nastrom1983} that helped motivate acceptance of the transition paradigm. 

Crucially, the implication is that any observed large-scale isobaric Hurst exponent cannot be simply assumed to be equal to the large-scale isoheight Hurst exponent, even though isobars are nearly flat. The more general lesson is that $H_h>1/3$ can be observed in the large-scale horizontal structure function if the observations depart even slightly from lying on a perfect isoheight surface.

Although our isoheight measurements are not isobaric, there is reason to believe that they also do not represent perfectly horizontal separations, mainly because radiosonde measurements are smoothed over a characteristic vertical distance of order $\sim 200\,\mathrm{m}$.\footnote{A second reason measurements may not represent perfect isoheights is measurement error in the vertical location, which would also tend to ``smooth'' the mean statistics in a similar manner. But given that measurement uncertainty in the GPS radiosondes is of order $\sim 10\,\mathrm{m}$, we assume vertical smoothing due to sonde inertia is more important.} Smoothing implies that a wind measurement at a given height is a function of the wind at nearby heights, plausibly causing vertically-separated fluctuations to influence any calculated horizontally-separated fluctuations. As a note, such smoothing cannot necessarily be removed by simply reprocessing the data. All sondes have a finite timescale of adjustment to the local wind in the presence of vertical wind shear. Such a timescale introduces an effective smoothing regardless of how the data are processed. 

To investigate the effect of vertical smoothing on horizontal statistics, we performed an experiment in a numerical hydrodynamic simulation using the System for Atmospheric Modeling \citep{khairoutdinov2003}. The simulation is of a tropical atmosphere and uses a numerical grid with $100\,\mathrm{m}$ spacing along all three directions in the layer where we performed our analysis, which was between $2$ and $10\,\mathrm{km}$. The domain size is $204\,\mathrm{km}\times 204\,\mathrm{km}$. Details of the simulation are provided in \citet{dazlichGigaLarge2013}.

\begin{figure}
    \includegraphics{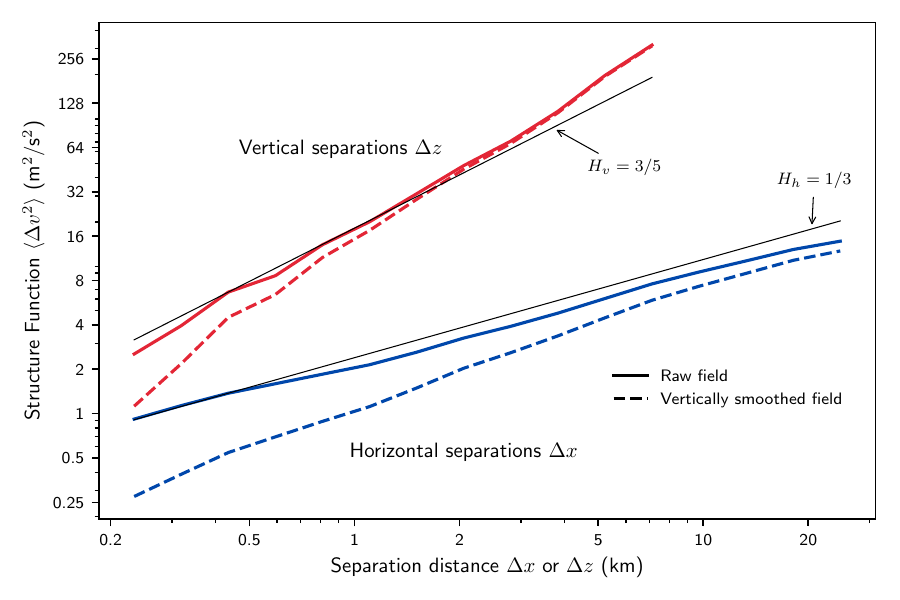}
    \centering \caption{Vertical (red) and horizontal (blue) structure functions calculated for the SAM simulation. Structure functions are calculated both for the original wind field with no smoothing applied (solid) and after the wind field is smoothed along the vertical direction (dashed). The theoretical values $H_h=1/3$ and $H_v=3/5$ according to Lovejoy-Schertzer theory are shown for reference (thin black).}\label{fig:SAM spectra}
\end{figure}

Vertical and horizontal structure functions calculated from the SAM wind field lie close to the Lovejoy-Schertzer scaling (Fig. \ref{fig:SAM spectra}). Omitting separation distances equal to one grid point, which may be more affected by numerical artifacts, calculated Hurst exponents for the raw wind field are $H_h = 0.305 \pm 0.008$ and $H_v=0.69 \pm 0.02$. As anticipated, when vertical smoothing is applied using a vertical Gaussian convolution with a standard deviation of two grid points or $200\,\mathrm{m}$, the horizontal Hurst exponent increases to $H_h=0.42 \pm 0.01$. The vertical Hurst exponent is also increased to a value of $H_v=0.79 \pm 0.03$. Notably, smoothing is only performed along the vertical direction but the horizontal Hurst exponent is nonetheless strongly affected.

Because the domain size is smaller than the horizontal measurements from IGRA, the result that vertical smoothing increases the horizontal Hurst  exponent should be viewed as qualitative rather than quantitative. More rigorous testing, perhaps with large-scale multifractal simulations with known Hurst exponents, would be necessary to determine whether the magnitude of the discrepancy between the IGRA-derived horizontal Hurst exponents and theory could be explained by vertical smoothing. Even so, Fig. \ref{fig:SAM spectra} suggests that vertical smoothing may plausibly explain why observations of $H_h$ are a little higher than expected by Lovejoy-Schertzer turbulence -- without needing to invoke any new set of physics such as quasi-geostrophic turbulence.

We should note that applying a vertical smoothing directly to Eqn. \ref{eq:2D structure function} can only decrease, rather than increase, the calculated horizontal Hurst exponent $H_h$. The reason is that Eqn. \ref{eq:2D structure function} applies for mean values of $\Delta v^2$ that are averaged over many realizations of the flow. This implies that the full distribution of values for $\Delta v^2$, rather than simply the mean $\langle \Delta v^2 \rangle$, must be considered to fully explain why $H_h$ is higher due to vertical smoothing. This is in contrast to the isobaric mechanism explanation for bias in measurements of $H_h$ suggested by \citet{lovejoy2009}, where the increase of $H_h$ may be derived directly from the mean statistics for $\Delta v^2$ represented by Eqn. \ref{eq:2D structure function} using the argument shown above. 

That the sounding data includes vertical smoothing may also explain \citet{lovejoy2007}'s prior finding of an increase in $H_v$ with altitude. Given that dropsondes such as those used by \citet{lovejoy2007} have a faster fall speed in the upper troposphere \citep[e.g.][]{vomel2023}, a given dropsonde-wind shear adjustment timescale would imply upper-tropospheric measurements are effectively smoothed over a larger vertical spatial scale than  lower-tropospheric measurements. If Hurst exponent calculations include scales affected by smoothing, a spurious altitude dependence of $H_v$ could result. \citet{lovejoy2007} included separations down to $5\,\mathrm{m}$ in their analyses, which is much smaller than our $200\,\mathrm{m}$ threshold and plausibly introduced a spurious dependence of $H_v$ on altitude that is not seen in Fig. \ref{fig:exponents with height}.

\section{Conclusions}

It is widely assumed that the dynamics of the atmosphere are controlled by a hierarchy of distinct dynamical mechanisms, each restricted to some limited range of spatial scales. The distribution of kinetic energy is thought to be determined by a quasi-two-dimensional enstrophy cascade at the largest scales \citep{charney1971,nastrom1984}, gravity waves at the mesoscale \citep{dewan1997,lindborg2006}, and a three-dimensional turbulent energy cascade at the smallest scales \citep{tennekes1972}. Such a hierarchy would imply clear transitions in the Hurst exponents for kinetic energy structure functions when calculated along either the horizontal or vertical directions, from $H=1$ at large scales to $H = 1/3$ at small scales. 

Here, we use high-resolution dropsonde and radiosonde measurements to calculate structure functions for horizontal wind separated both horizontally and vertically. We find a Hurst exponent close to $H_v\approx 0.6$ for vertical separations between $200\,\mathrm{m}$ and $8\,\mathrm{km}$, which is inconsistent with both small-scale isotropic turbulence and mesoscale gravity waves. Along the horizontal direction, large scale structure functions show a Hurst exponent with value $H_h\approx 0.4$ 
for separation scales ranging from  $200\,\mathrm{km}$ to $2000\,\mathrm{km}$, which is inconsistent with a large-scale enstrophy cascade.  We argue that these measured structure functions  are closely consistent with a lesser known theory of ``Lovejoy-Schertzer'' turbulence with $H_h = 1/3$ and $H_v = 3/5$ at all scales \citep{schertzer1985}. We show that the small difference between the observed and Lovejoy-Schertzer value of $H_h$ is plausibly due to vertical smoothing of radiosonde data. 

Thus, we find that the canonical ``transition'' paradigm has little empirical support. Instead, it appears that the dynamics of the troposphere and most of the stratosphere are controlled by a single wide-ranging anisotropic turbulent cascade rather than a hierarchy of independent dynamical mechanisms.  Looking forward, more measurements of structure functions or spectra as a function of separation direction will be necessary to confirm Lovejoy-Schertzer scaling. Scales smaller than $200\,\mathrm{m}$, which were not resolved here, are of particular interest for determining whether small-scale turbulence is isotropic with $H_h = H_v$. Simultaneous wind observations separated in both the horizontal and vertical direction will be necessary for such an analysis. The extent to which vertical smoothing affects calculations of horizontally-separated structure functions could also be quantified, perhaps by considering multifractal simulations with known exponents \citep{lovejoy2010}.

\appendix
\section{Vertical Hurst exponents for individual meteorological stations}  \label{sec:individual stations}  

Reported vertical Hurst exponents from the IGRA data were computed after the structure function was averaged over all available stations to obtain a global mean value of $H_v$. Here, we calculate the exponents for individual stations by identifying which stations contained at least 730 complete soundings, corresponding to at least two years. Then, mean structure functions and Hurst exponents were calculated for individual $2\,\mathrm{km}$-thick vertical layers as in Sect. \ref{sec:results} and Fig. \ref{fig:exponents with height}. Finally, as visual inspection of plotted structure functions was not possible, we filtered the calculated exponents by requiring the 95\% confidence interval of the least-squares fit to be less than 0.05. This ensures that we are not assuming a power-law form for structure functions that are poorly described by a power-law fit. This filtering process was only used in this section and resulted in a total of 184 stations for consideration.

\begin{figure}
    \includegraphics{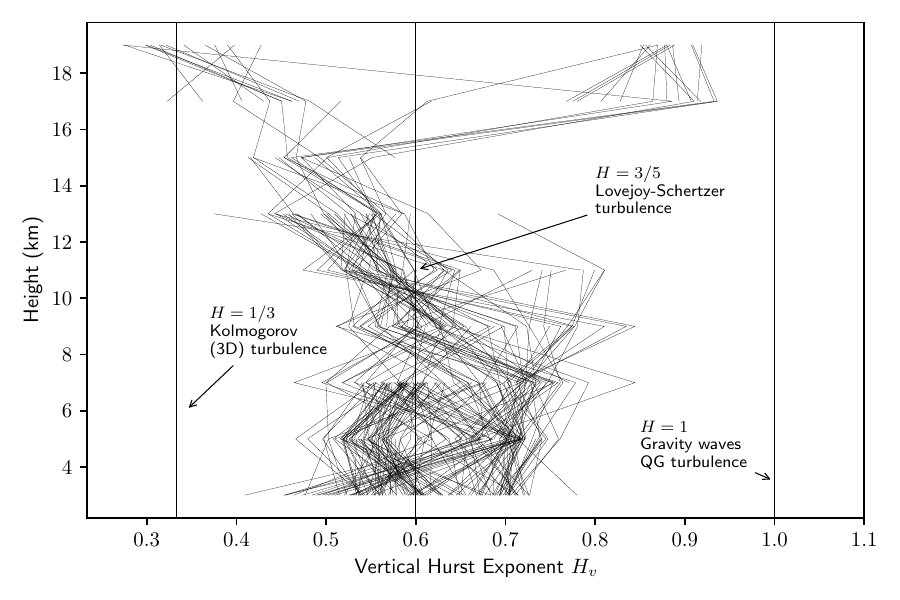}
    \centering \caption{As in Fig. \ref{fig:exponents with height}, but for individual IGRA stations (thin black lines).  }\label{fig:exponents with height each station}
\end{figure}

Figure \ref{fig:exponents with height each station} shows the vertical Hurst exponents as a function of height for each IGRA station. Overall, a large majority of values are much closer to the Bolgiano-Obukhov value of 3/5 than the values of 1/3 or 1 that correspond to three-dimensional or two-dimensional turbulence, respectively. There also appears to be a bimodal distribution below $8\,\mathrm{km}$, with a cluster of values near 0.6 and a cluster closer to 0.75. At first glance, this might indicate evidence of two separate turbulence regimes, possibly depending on latitude or some other climatological factor. However, we also observe a nearly perfect split between the two regimes based on sounding nationality. For example, for the layer between $4\,\mathrm{km}$ and $6\,\mathrm{km}$, 82 out of 84 U.S.A.-based soundings had a Hurst exponent smaller than 0.63, whereas 79 out of 84 Hurst exponents measured from China-based soundings were larger than 0.63. Recall that the value of the exponent is sensitive to data processing methods, such as the period of the vertical smoothing applied to the raw data as shown in Sect. \ref{sec:discussion}. Since it is likely that the methods are uniform within a country but differ between different countries, the bimodal distribution in Fig. \ref{fig:exponents with height each station} is likely an indication of data processing artifacts rather than some dependence on local meteorology.

\section*{Acknowledgments}

This research has been supported by the National Science Foundation (grant no. PDM-2210179).

\section*{Data Availability}
All datasets used in this study are publicly archived \citep{activatedataset,igradataset,noaahurricanedataset}.

%
\bibliographystyle{plainnat}  
\bibliography{sources.bib}





\end{document}


\renewcommand{\thesection}{S\arabic{section}}
\renewcommand{\thefigure}{S\arabic{figure}}
\renewcommand{\thetable}{S\arabic{table}}
\renewcommand{\theequation}{S\arabic{equation}}

\title{Supplement to: Global sonde datasets do not support a mesoscale transition in the turbulent energy cascade}
\author{Thomas D. DeWitt}
\date{October 23, 2025}
\maketitle

\section{A generalized anisotropic structure function}

\begin{figure}
    \centering
    \includegraphics[width=.3\linewidth]{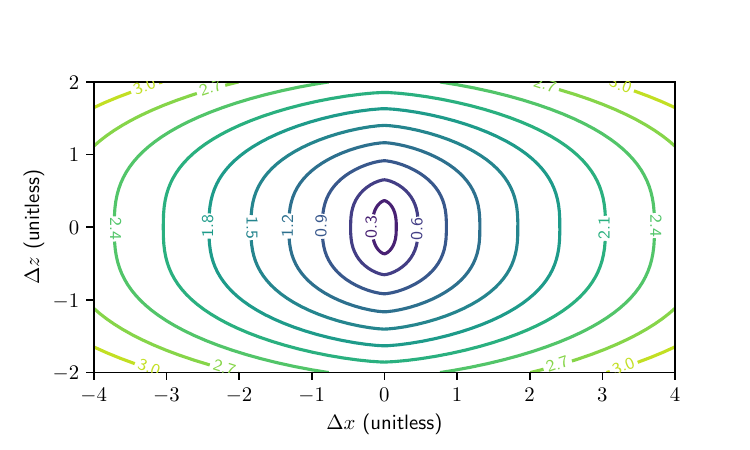}
    \includegraphics[width=0.3\linewidth]{figures/Example_2D_structure_function}
    \includegraphics[width=0.3\linewidth]{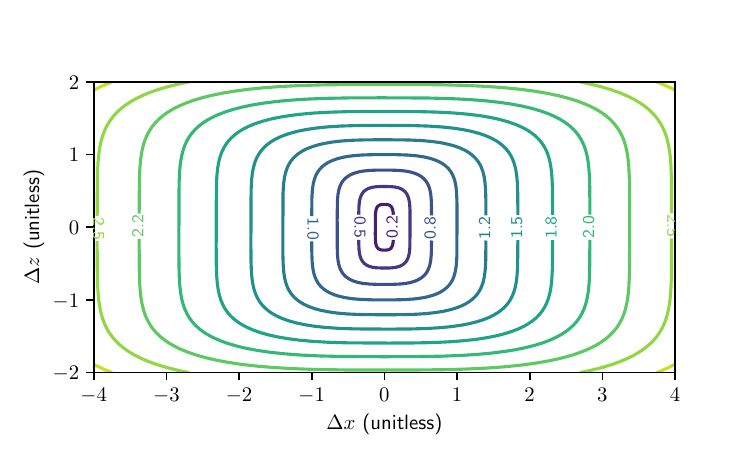}
    \caption{Contour plot of dimensionless form of Eqn. \ref{eq:2D structure function general}, as in Fig. 2 but for $\eta=0.75$, $\eta=1$, and $\eta=1.5$. The value $\eta=1$ was used in the main text.}\label{fig:examples}
\end{figure}

The two-dimensional structure function represented by Eqn. 9 in the main text is not the unique choice if we simply require Eqns. 7 to be recovered when $\Delta x = 0$ or $\Delta z=0$. To obtain Eqn. 9, we added the requirement that isotropic turbulence is recovered when $H_h=H_v$ and $\varphi_v=\varphi_h$. Here we drop this assumption by introducing an additional parameter $\eta$ such that
\begin{align}
    \langle \Delta v(\Delta x, \Delta z)^2\rangle  = \left(\varphi_h^{\eta/H_h} \Delta x^{2\eta}  +\varphi_v^{\eta/H_h} \Delta z^{H_v\eta/H_h}\right)^{H_h/\eta}. \label{eq:2D structure function general}
\end{align}
When $\eta=1$, Eqn. 9 in the main text is recovered. Otherwise, Eqns. 7 are still obtained when $\Delta x = 0$ or $\Delta z=0$, but isotropic turbulence cannot be recovered.
Visually, the parameter $\eta$ makes the isolines of $\Delta v$ more or less ``boxy,'' as shown in Fig. \ref{fig:examples}. 

\begin{figure}
    \centering
    \includegraphics{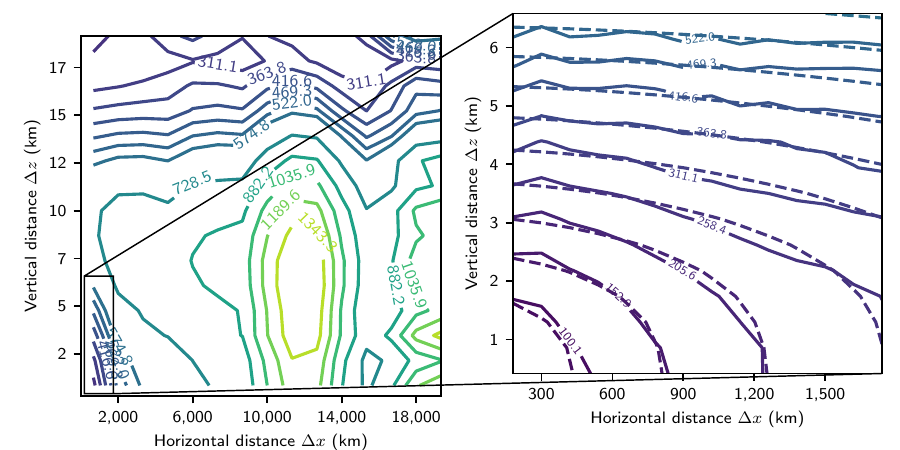}
    \caption{As in Fig. 7, but using Eqn. \ref{eq:2D structure function general} for the fit.}\label{fig:2D general SF}
\end{figure}

Fitting Eqn. \ref{eq:2D structure function general}, rather than Eqn. 9, to the empirical structure function (Fig. \ref{fig:2D general SF}) resulted in best-fit values of $\phi_h=0.015 \pm 0.007, \phi_v=0.006 \pm 0.001, H_h = 0.34 \pm 0.02, H_v = 0.65 \pm 0.01$, and $\eta=0.81 \pm 0.07$, which are close to the values reported in the main text for $\eta=1$. The contour lines of the fit in Fig. \ref{fig:2D general SF} also provide a slightly better match to the empirical contour lines as compared to Fig. 7.

\clearpage
\section{First and third order structure functions}

\begin{figure}
    \centering
    \includegraphics{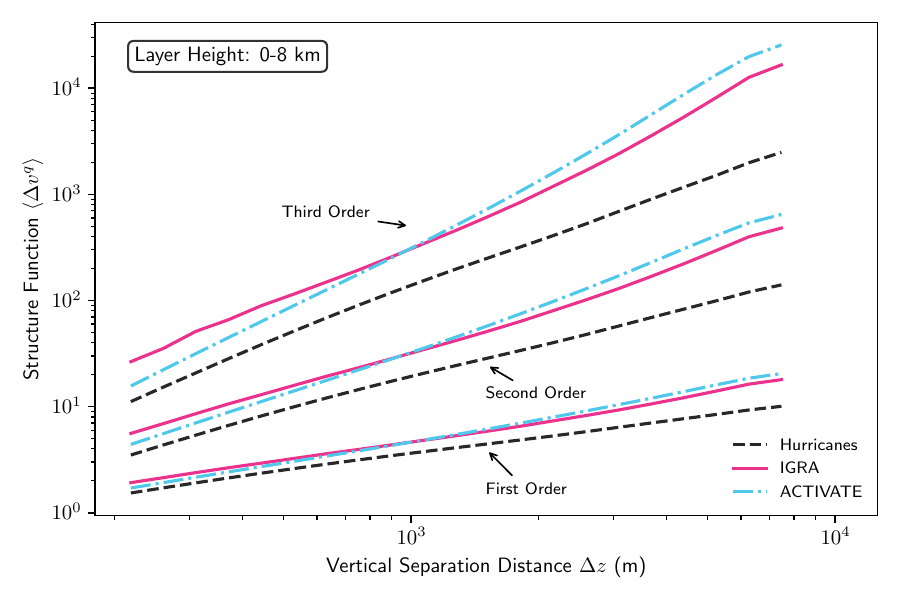}
    \caption{As in Fig. 4, but for first-, second-, and third-order structure functions $\langle \Delta v^2\rangle$ (lower), $\langle \Delta v^2\rangle$ (middle) and $\langle \Delta v^3\rangle$ (upper).  }\label{fig:8km vertical spectrum order 2 and 3}
\end{figure}

\begin{figure}
    \centering
    \includegraphics{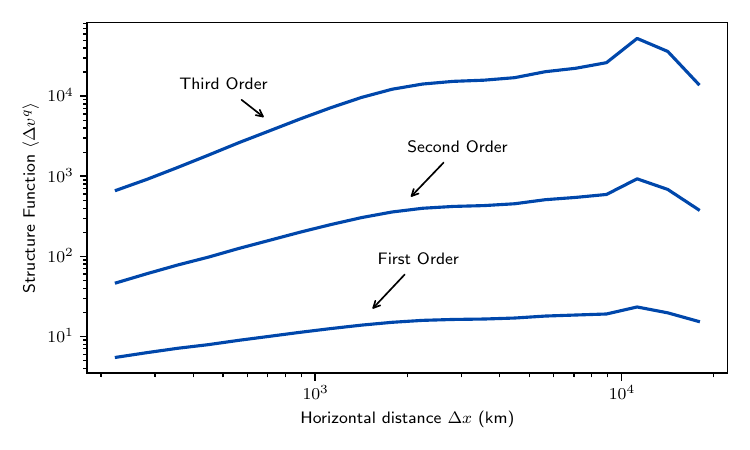}
    \caption{As in Fig. 6, but for first-, second-, and third-order structure functions $\langle \Delta v^2\rangle$ (lower), $\langle \Delta v^2\rangle$ (middle) and $\langle \Delta v^3\rangle$ (upper).  }\label{fig:horizontal structure function order 2 and 3}
\end{figure}

\begin{figure}
    \centering
    \includegraphics{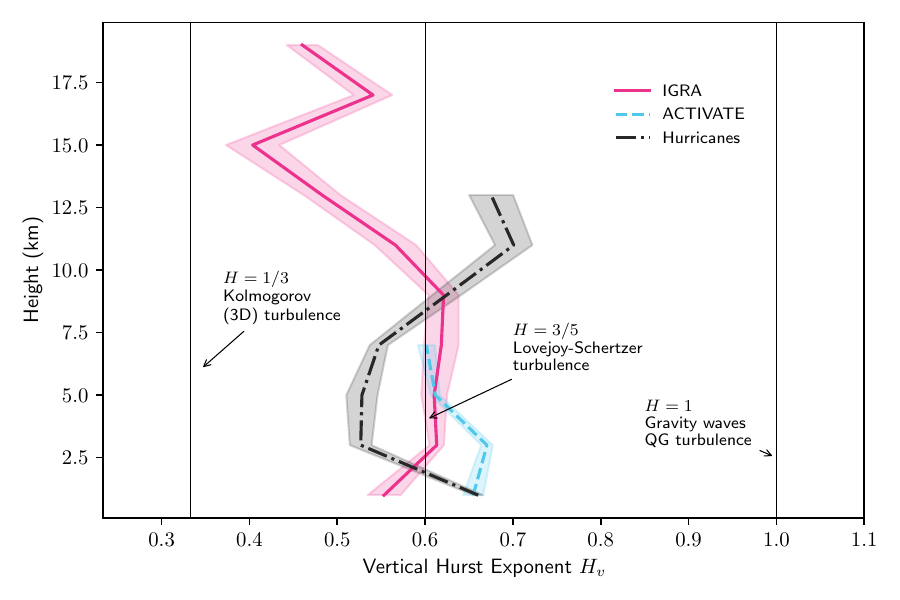}
    \caption{As in Fig. 5, but for first-order structure functions $\langle \Delta v\rangle$.  }\label{fig:exponents with height higher order 2}
\end{figure}

\begin{figure}
    \centering
    \includegraphics{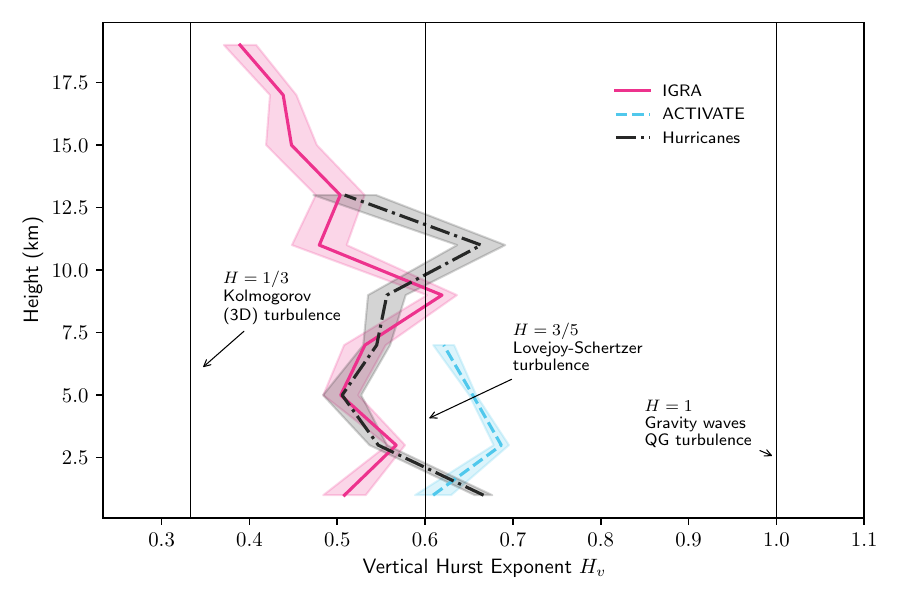}
    \caption{As in Fig. 5, but for third-order structure functions $\langle \Delta v^3\rangle$.  }\label{fig:exponents with height higher order 3}
\end{figure}

In the main text we prefer second-order structure functions of velocity increments, which represent kinetic energy $\Delta v^2$, because they are most commonly examined in prior studies of turbulence. However, structure functions may be defined for arbitrary orders $q$ as
\begin{align}
    \langle \Delta v ^q\rangle \sim \Delta r ^{\zeta(q)}. \label{eq:order q structure function}
\end{align} 
For a nonintermittent turbulent cascade, the structure function exponent $\zeta(q)$ is simply equal to $qH$. For the more realistic intermittent case, in general $\zeta(q) < qH$ for $q>1$ \citep{lovejoy2013}. If the intermittency is weak, we can reasonably infer $H$ for structure functions with $q>1$ using the equation $H\approx\zeta(q)/q$ with the expectation that this method may slightly underestimate $H$. The main text used this method for $q=2$. In Figs. \ref{fig:8km vertical spectrum order 2 and 3}-\ref{fig:exponents with height higher order 3} we reproduce the Figs. 4-8 for first order structure functions, where it is expected that $\zeta=H$ for a multiplicative cascade  \citep{lovejoy2013}, and for $q=3$ for comparison with other prior studies. Table \ref{tab:high_order_structure_function_exponents} lists the calculated exponents derived from structure functions calculated for $q=1$, $q=2$, and $q=3$.

\begin{table}[h]
\centering
\caption{Structure function exponents for orders $q=1,2,3$ (Eqn. \ref{eq:order q structure function}) and estimated values for the Hurst exponent. }
\label{tab:high_order_structure_function_exponents}
\begin{tabular}{|c|c|c|c|c|c|}
\hline
Dataset & Direction & Order ($q$) & Structure Function Exponent $\zeta(q)$ & Inferred $H=\zeta(q)/q$ \\
\hline
IGRA & Vertical & 1 & $0.62 \pm 0.02$ & $0.62 \pm 0.02$ \\
 & & 2 & $1.24 \pm 0.04$ & $0.62 \pm 0.02$ \\
 & & 3 & $1.80 \pm 0.06$ & $0.60 \pm 0.02$ \\
\hline
ACTIVATE & Vertical & 1 & $0.70 \pm 0.01$ & $0.70 \pm 0.01$ \\
 & & 2 & $1.43 \pm 0.03$ & $0.71 \pm 0.01$ \\
 & & 3 & $2.12 \pm 0.04$ & $0.71 \pm 0.01$ \\
\hline
Hurricanes & Vertical & 1 & $0.523 \pm 0.007$ & $0.523 \pm 0.007$ \\
 & & 2 & $1.03 \pm 0.02$ & $0.513 \pm 0.008$ \\
 & & 3 & $1.51 \pm 0.03$ & $0.504 \pm 0.009$ \\
\hline
IGRA & Horizontal & 1 & $0.49 \pm 0.02$ & $0.49 \pm 0.02$ \\
 & & 2 & $1.00 \pm 0.04$ & $0.50 \pm 0.02$ \\
 & & 3 & $1.44 \pm 0.04$ & $0.48 \pm 0.01$ \\
\hline
\end{tabular}
\end{table}


\clearpage
\section{Structure functions for individual 2\,km-thick layers}

\begin{figure}
    \centering
    \includegraphics{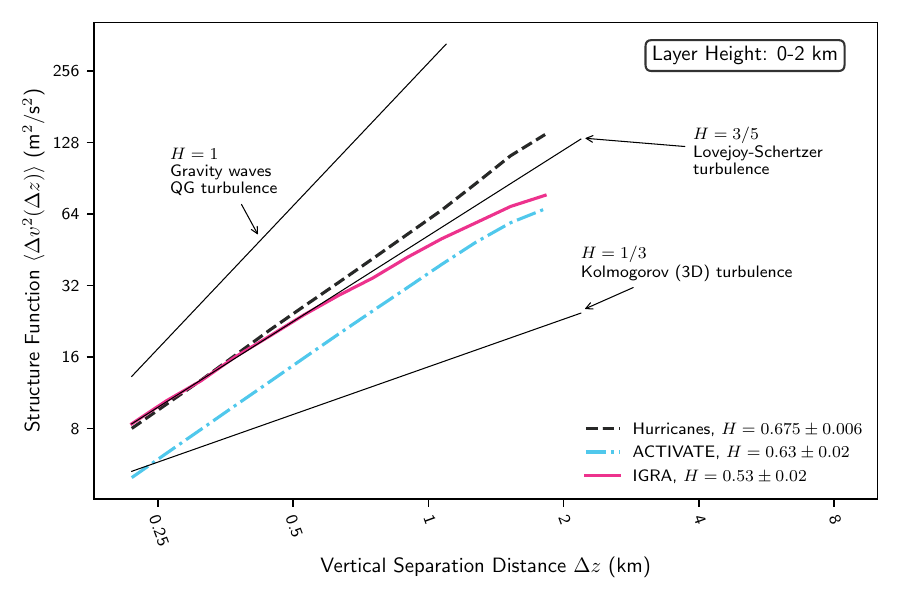}
    \caption{As in Fig. 4, but for altitudes between 0 and 2 km.}
\label{fig:vertical SF layer 0-2km}
\end{figure}

\begin{figure}
    \centering
    \includegraphics{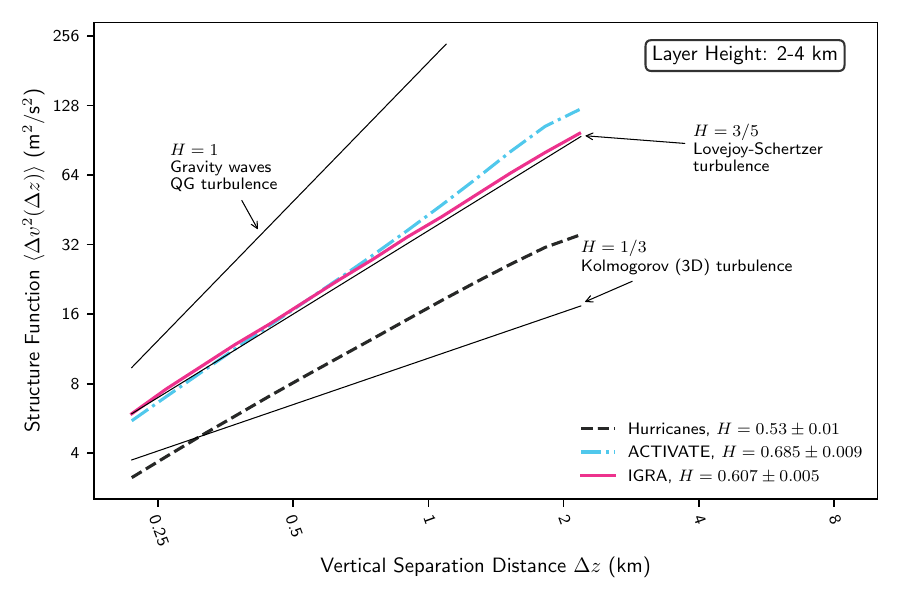}
    \caption{As in Fig. 4, but for altitudes between 2 and 4 km.}
\label{fig:vertical SF layer 2-4km}
\end{figure}

\begin{figure}
    \centering
    \includegraphics{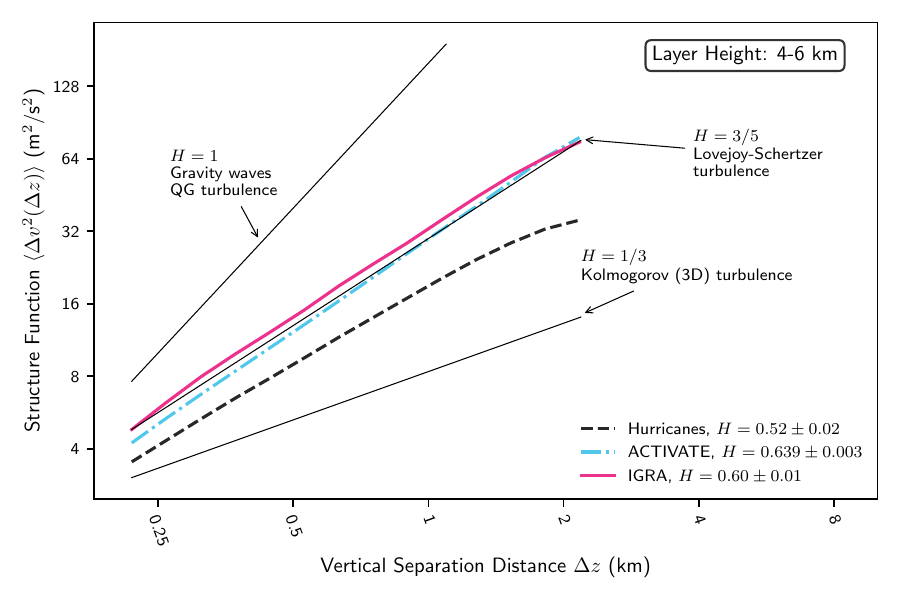}
    \caption{As in Fig. 4, but for altitudes between 4 and 6 km.}
\label{fig:vertical SF layer 4-6km}
\end{figure}

\begin{figure}
    \centering
    \includegraphics{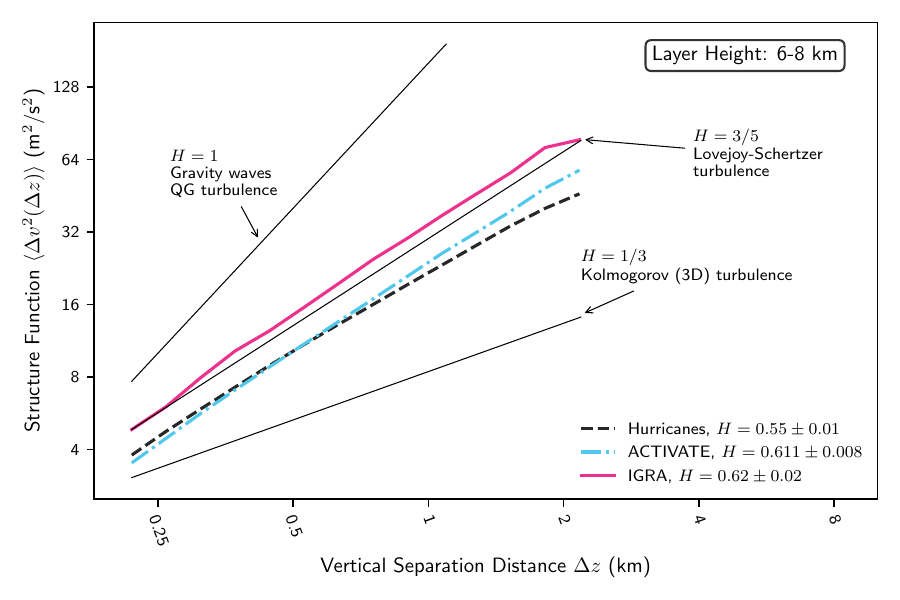}
    \caption{As in Fig. 4, but for altitudes between 6 and 8 km.}
\label{fig:vertical SF layer 6-8km}
\end{figure}

\begin{figure}
    \centering
    \includegraphics{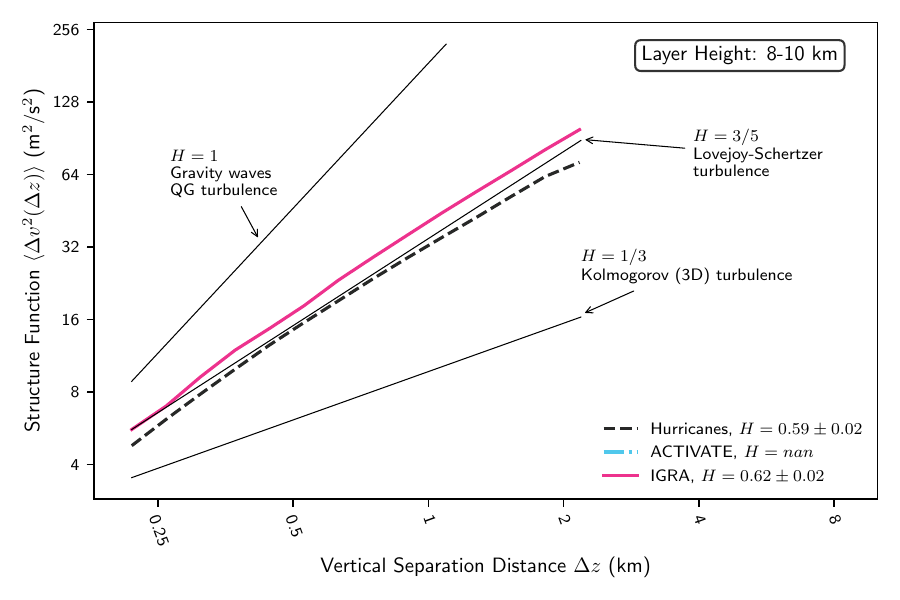}
    \caption{As in Fig. 4, but for altitudes between 8 and 10 km.}
\label{fig:vertical SF layer 8-10km}
\end{figure}

\begin{figure}
    \centering
    \includegraphics{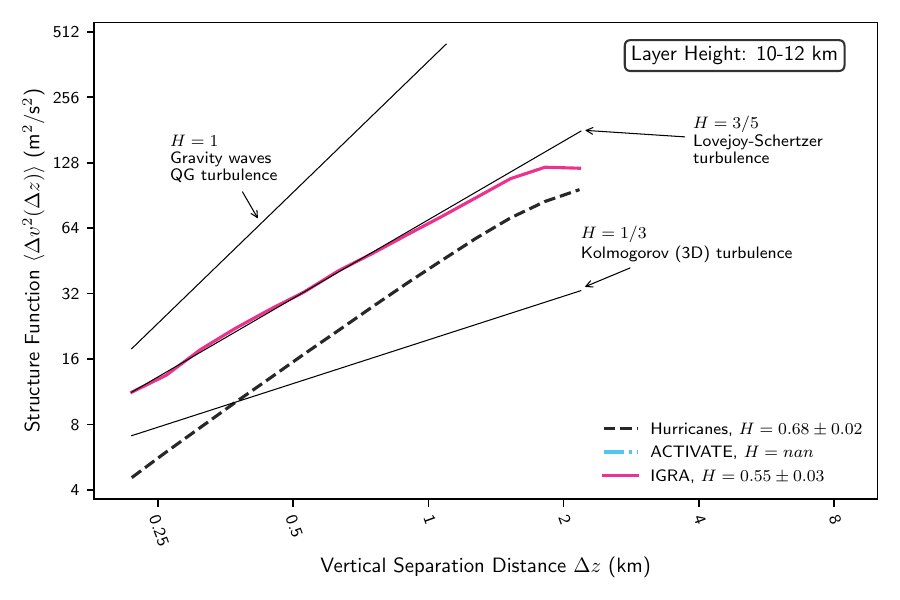}
    \caption{As in Fig. 4, but for altitudes between 10 and 12 km.}
\label{fig:vertical SF layer 10-12km}
\end{figure}

\begin{figure}
    \centering
    \includegraphics{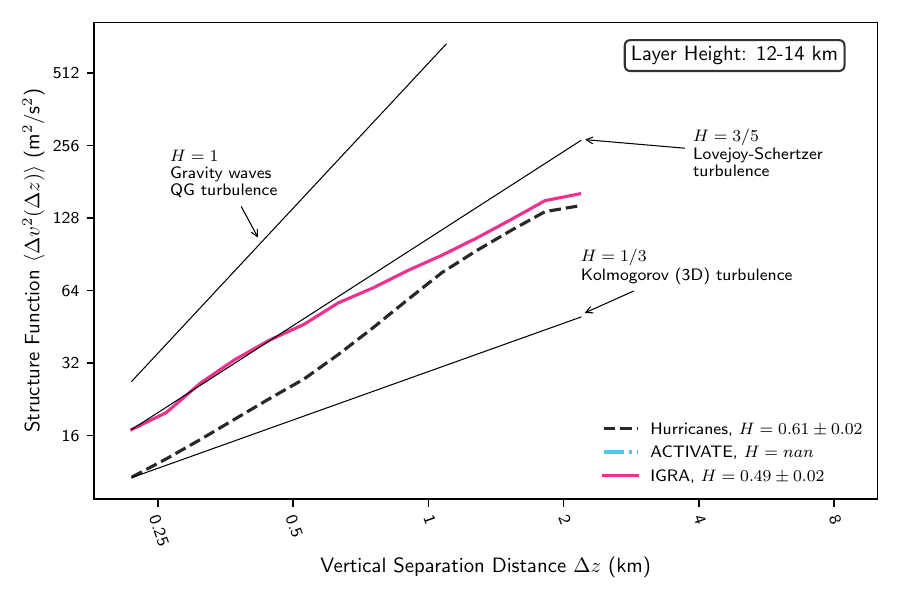}
    \caption{As in Fig. 4, but for altitudes between 12 and 14 km.}
\label{fig:vertical SF layer 12-14km}
\end{figure}

\begin{figure}
    \centering
    \includegraphics{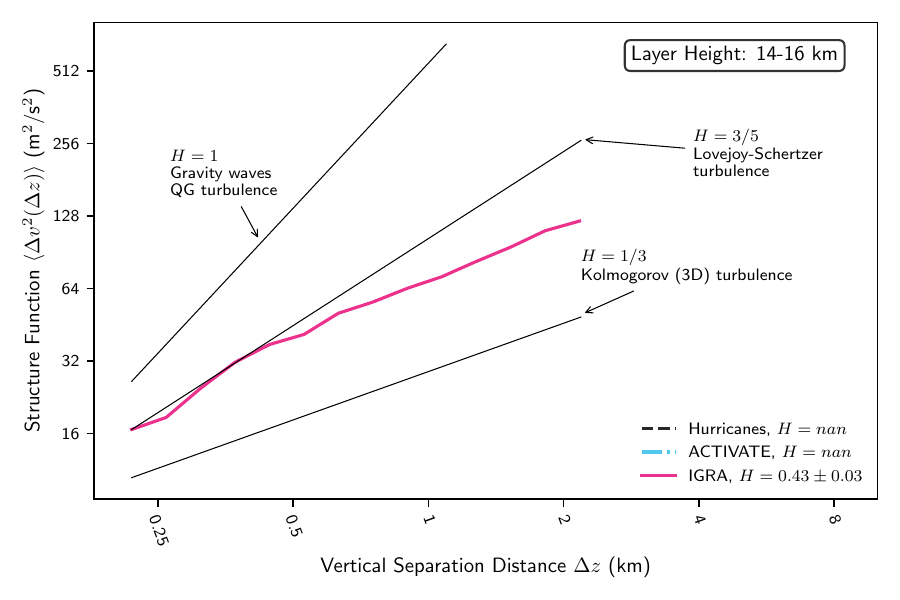}
    \caption{As in Fig. 4, but for altitudes between 14 and 16 km.}
\label{fig:vertical SF layer 14-16km}
\end{figure}

\begin{figure}
    \centering
    \includegraphics{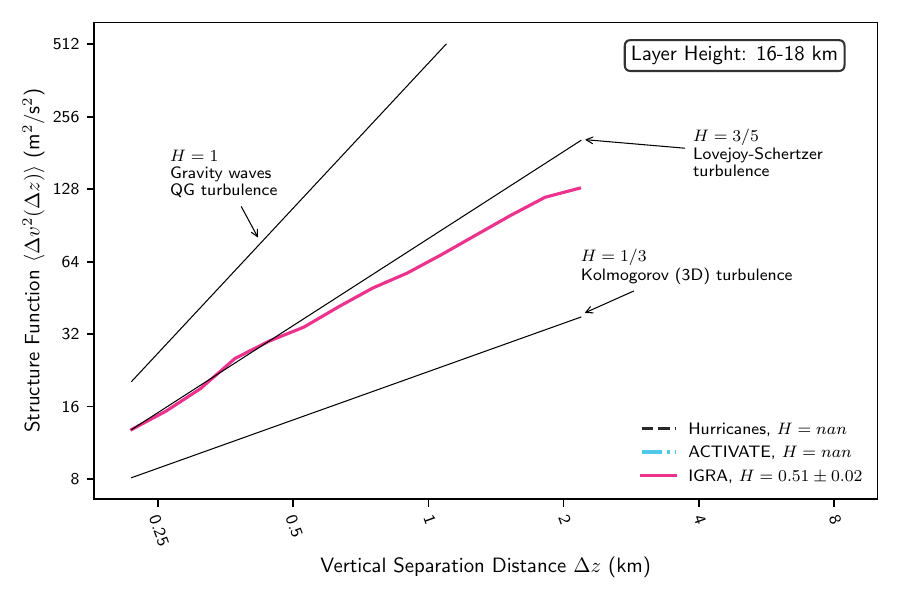}
    \caption{As in Fig. 4, but for altitudes between 16 and 18 km.}
\label{fig:vertical SF layer 16-18km}
\end{figure}

\begin{figure}
    \centering
    \includegraphics{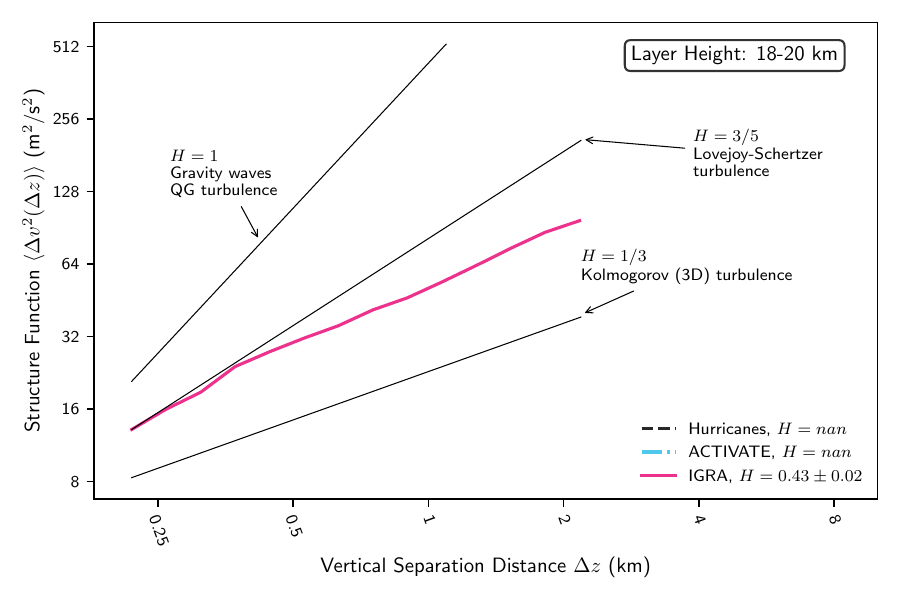}
    \caption{As in Fig. 4, but for altitudes between 18 and 20 km.}
\label{fig:vertical SF layer 18-20km}
\end{figure}

Figures \ref{fig:vertical SF layer 0-2km} to \ref{fig:vertical SF layer 18-20km} display the structure functions for each altitude layer that were used to calculate $H_v$ as a function of height in Fig. 5.

\clearpage
\section{Additional horizontal structure functions}

\begin{figure}
    \centering
    \includegraphics{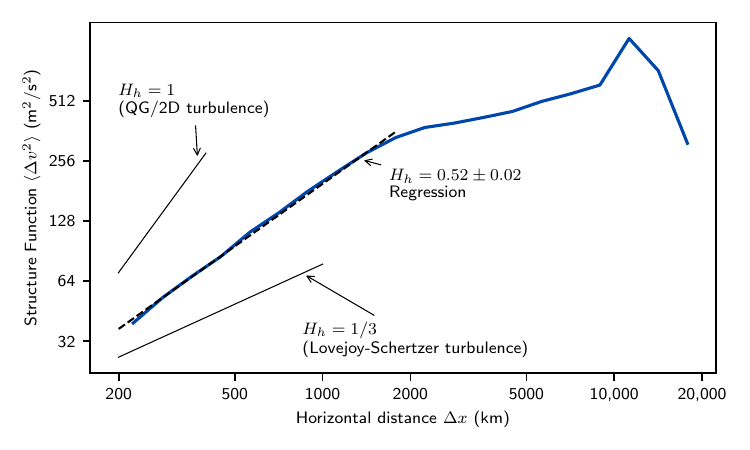}
    \caption{As in Fig. 6, but for observation pairs separated by at most 5 minutes and 5\,m in altitude. }
\label{fig:small sep}
\end{figure}

In this section we report four additional horizontal structure functions calculated from IGRA data. Figure \ref{fig:small sep} shows a horizontal structure function as in Fig. 6 but for observation pairs separated by at most 5 minutes and 5\,m in altitude as described in Section 3.1.

\begin{figure}
    \centering
    \includegraphics{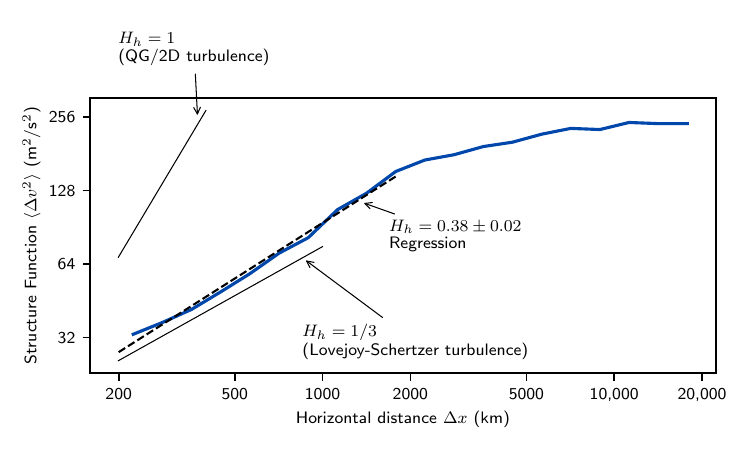}
    \caption{As in Fig. 6, but for observation pairs between -20° and 20° in latitude. }
\label{fig:tropics}
\end{figure}

\begin{figure}
    \centering
    \includegraphics{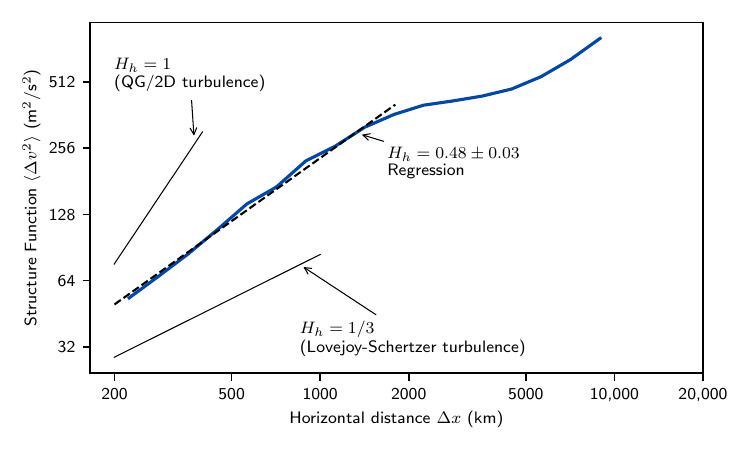}
    \caption{As in Fig. 6, but for observation pairs between 45° and 90° in latitude. }
\label{fig:highlat}
\end{figure}

Figure \ref{fig:tropics} shows a structure function calculated for observation pairs between -20° and 20° in latitude, while Fig. \ref{fig:highlat} shows a structure function calculated for observation pairs between 45° and 90° in latitude.

\begin{figure}
    \centering
    \includegraphics{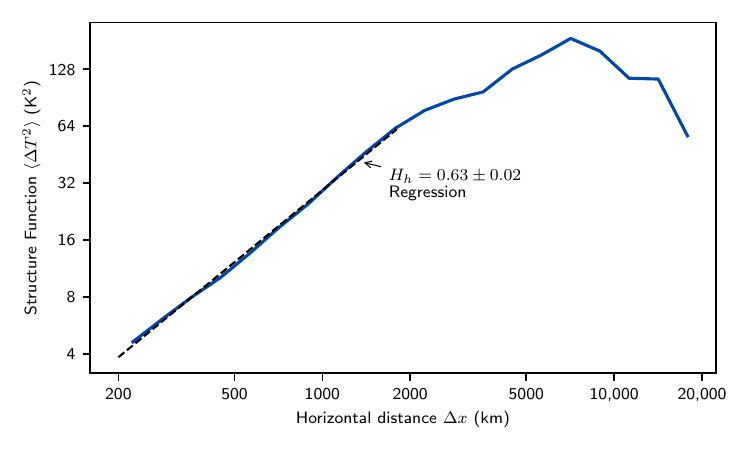}
    \caption{As in Fig. 6, but for temperature $T$ rather than kinetic energy, where $\Delta T^2\sim \Delta x^{2H_h}$.}
\label{fig:temp}
\end{figure}

\begin{figure}
    \centering
    \includegraphics{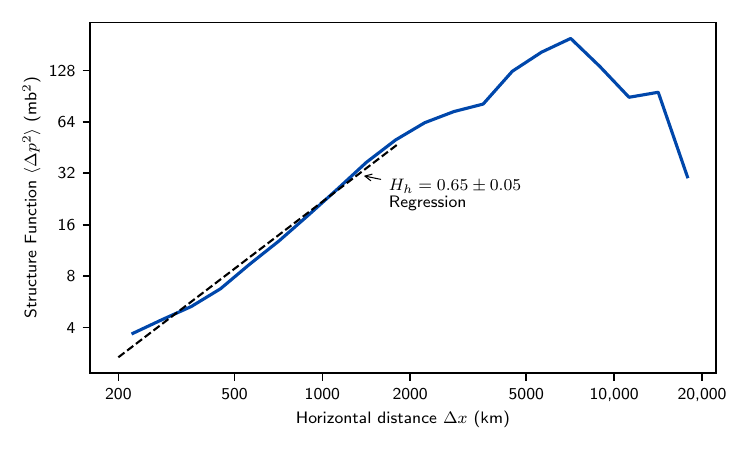}
    \caption{As in Fig. 6, but for pressure $p$ rather than kinetic energy, where $\Delta p^2\sim \Delta x^{2H_h}$.}
\label{fig:pressure}
\end{figure}

Figures \ref{fig:temp} and \ref{fig:pressure} show horizontal structure functions for temperature variance $T^2$ and pressure variance $p^2$, respectively, rather than kinetic energy.

\clearpage

\bibliographystyle{plainnat}
\bibliography{sources.bib}